\documentclass[referee,usenatbib]{mnras}
\usepackage{newtxtext,newtxmath}

\usepackage[T1]{fontenc}
\DeclareRobustCommand{\VAN}[3]{#2}
\let\VANthebibliography\thebibliography
\def\thebibliography{\DeclareRobustCommand{\VAN}[3]{##3}\VANthebibliography}

\usepackage{graphicx}	
\usepackage{amsmath}	
\usepackage{amssymb}	

\title[Chandra Spectroscopy of Supernova Remnant DEM L71]{Spatially Resolved Chandra Spectroscopy of Supernova Remnant DEM L71 in the Large Magellanic Cloud}

\author[Alan \&  Bilir]{
N. Alan,$^{1}$\thanks{E-mail: neslihan.aln@gmail.com}
and
S. Bilir,$^{1}$
\\
$^1$Istanbul University, Faculty of Science, Department of Astronomy and Space
Sciences, 34119, Istanbul, Turkey\\
}
\date{Accepted XXX. Received YYY; in original form ZZZ}

\pubyear{2015}

\begin{document}
\label{firstpage}
\pagerange{\pageref{firstpage}--\pageref{lastpage}}
\maketitle

\begin{abstract}
We present a detailed X-ray spectroscopic study of the supernova remnant DEM L71 in the Large Magellanic Cloud. Based on deep $\sim 103$ ks archival {\it Chandra} data, we perform a detailed spatially resolved spectral analysis of DEM L71. We analyze regional spectra extracted from thin-sliced regions along several different azimuthal directions of the SNR to construct radial profiles of elemental abundances for O, Ne, Mg, Si, and Fe. Our elemental abundance measurements reveal an asymmetrical spatial distribution of metal-rich ejecta gas. Especially the asymmetry on the western part of the central Fe distribution is remarkable. While the location of the contact discontinuity is generally at $\sim 5$ pc from the geometric center of the X-ray emission of DEM L71, it is uncertain in western part of the remnant. Fe is enhanced in the ejecta while O and Ne abundances are generally negligible. This finding confirms the Type Ia origin of DEM L71. We estimate an upper limit on the Sedov age of $\sim 6,660\pm 770$ yr and explosion energy of $\sim 1.74\pm 0.35\times 10^{51}$ erg for the remnant. This explosion energy estimate is consistent with a canonical explosion of a Type Ia supernova remnant.
\end{abstract}




\section{Introduction}

Supernova (SN) explosions, the most energetic stellar events known, play an important role in shaping the energy density, chemical enrichment, and evolution of galaxies. The ejecta from supernovae seed the galaxies with heavy elements (e.g., carbon, oxygen, neon, iron, and higher atomic numbers) and the resulting elemental distribution and composition can be used as clues to understand the nature of the supernova progenitors. 

Type Ia supernovae are the thermonuclear explosions of carbon oxygen (CO) white dwarfs within close binary systems. When a CO white dwarf approaches the Chandrasekhar limit through mass accretion from a non-degenerate companion star \citep[single-degenerate (SD);][]{Whelan73, Nomoto82}, it becomes unstable and a thermonuclear runaway occurs. A thermonuclear explosion may also be produced by the merging of two white dwarfs \citep[double-degenerate (DD);][]{Iben84}. The detailed physics involved in these thermonuclear explosions and the nature of their progenitor systems are under debate, and several models exist to describe Type Ia SN explosion mechanisms \citep[e.g.][]{Wang12, Maoz14, Ruiz-Lapuente14, Ruiz-Lapuente18, Wang18}.

The supernova remnant (SNR) DEM L71 in the Large Magellanic Cloud (LMC) was discovered in the optical band by \citet{Davies76}. The first X-ray observations of DEM L71 were obtained by {\it Einstein} \citep{Long81}. Since then, it has been further observed in X-rays using {\it ASCA} \citep{Hughes98}, {\it Chandra} \citep{Hughes03, Rakowski03}, and {\it XMM-Newton} \citep{vanderHeyden03,  Maggi16}. {\it ASCA} observations indicated that the remnant were enhanced in Fe \citep{Hughes98}, which is generally considered to be a characteristic feature from the thermonuclear explosion of a CO white dwarf. This was further confirmed by subsequent high-resolution observations with {\it Chandra} \citep{Hughes03, Rakowski03} and {\it XMM-Newton} \citep{vanderHeyden03}. These observations, along with associated Fe ejecta mass estimates of about $1.4 M_{\odot}$ \citep{Hughes03, vanderHeyden03}, suggest that DEM L71 is the remnant of a Type Ia SN. Optical observations of the Balmer-dominated shock velocities by \citet{Ghavamian03} yield an age estimate of $4,360\pm90$ yr. \citet{Pagnotta15} utilized optical images ($g'$, $r'$, $i'$, and $\rm{H}_{\alpha}$) to identify a great number of possible candidates for the SNs surviving companion; but it has not been detected. Recently, \citet{Siegel20} analyzed the {\it XMM-Newton} data of the DEM L71 and estimated the total mass of the swept-up interstellar medium (ISM) as $228\pm 23 M_{\odot}$.

In previous X-ray studies, several sub-regions characteristically representing emission from the swept-up ambient gas and the shocked metal-rich ejecta gas were examined, but an extensive X-ray study of the entire remnant has not yet been performed. In this work, based on the archival {\it Chandra} data, we perform the spatially-resolved spectral analysis of the entire SNR to provide unprecedented details on radial and azimuthal structures of DEM L71. We describe the X-ray data and the data reduction in Section 2. We present the X-ray imaging and spectral analysis of the SNR in Section 3. We discuss the results and compare this work to other abundance measurements in Section 4.

\section{X-ray data and reduction}

We used two X-ray observations ($\sim 103$ ks in total) of SNR DEM L71 obtained with the S3 chip in {\it Chandra}'s Advanced Charged Couple Device Imaging Spectrometer (ACIS-S) detector array \citep{Bautz98}. We summarize these {\it Chandra} observations in Table 1. 

We reprocessed each individual ObsID with CIAO version 4.9 via the \texttt{chandra\_repro} script. We removed time intervals that shows high particle background fluxes (by a factor of $\sim2$ higher than the mean background). After the data reduction, the total effective exposure time is $\sim 100$ ks.

\begin{table}
  \centering
  \caption{Observation log for Archival {\it Chandra} ACIS Data of DEM L71.}
    \begin{tabular}{cccc}
    \hline
    ObsID           & Date              & Exposure (ks)   & Detector \\
    \hline
    3876            & 2003 July 4       & 48.98           & ACIS-S \\
    4440            & 2003 July 6       & 54.27           & ACIS-S \\
    \hline
    \end{tabular}%
\end{table}%

\section{Analysis and Results}
\subsection{Imaging}
Figure 1 represents an ACIS-S3 broadband and an X-ray three-colour (RGB) images of DEM L71. To create these images we combined all individual archival {\it Chandra} data. The X-ray morphology of the DEM L71 shows a bright outer rim and a faint, diffuse central nebula, usually interpreted as the forward-shocked ambient medium and reverse-shocked metal-rich ejecta, respectively (Figure 1a). The X-ray emission of DEM L71 shows broadly similar intrinsic emissivities around the outer rim \citep{Hughes03}. There are no large variations in abundance or thermodynamic state along these outer rim regions \citep{Rakowski03}, so that surface brightness variations largely trace variations in the ambient gas density around DEM L71. In general, the main outermost boundary shows a slightly elliptical (or nearly circular) morphology. As shown in Figure 1, in north-south the outer rim spans 83$''$ ($\sim 20$ pc, the distance of LMC was assumed as 50 kpc \citep{Freedman01}) while in the east-west direction it is 60$''$ ($\sim 14$ pc) across. In Figure 1b, we show the {\it Chandra} RGB image with red, green and blue corresponding to the  soft (0.3-0.7 keV),  medium  (0.7-1.1 keV), and hard (1.1-4.0 keV) energy bands, respectively. The energy bands displayed in each colour were chosen to emphasize major atomic line emission that illustrates the distribution of electron temperatures and ionization states across DEM L71. We used the sub-band images with the native pixel scale of the ACIS detector ($0.''49$ pixel$^{-1}$) and then adaptively smoothed them. The region between the bright outer rim and the inner ejecta nebula of the remnant show harder X-ray emission (blue), especially the east part, however X-ray emission in the outermost boundary is generally soft (red). While the innards X-ray emission from the shocked ejecta is faded, the broadband intensity peaks in brightness at east and west rim of the remnant. This intensity peak corresponds to soft (red) parts in the three-colour image. 

\begin{figure}
\begin{center}
\includegraphics[scale=0.50]{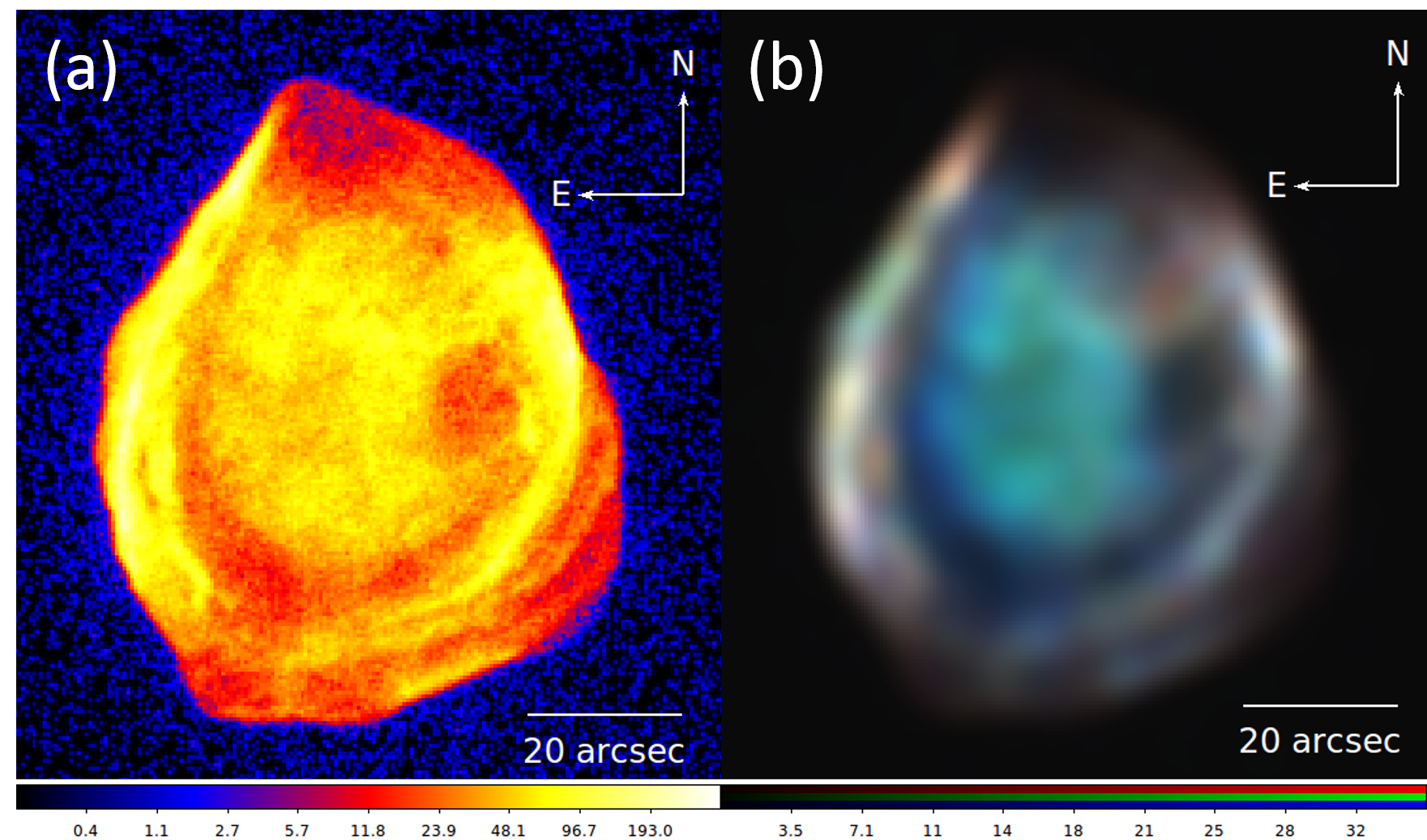}
\caption{(a) The broadband ACIS image of DEM L71 (0.3-4.0 keV). (b) The three-colour image of DEM L71: Red = 0.3-0.7 keV, green = 0.7-1.1 keV, and blue = 1.1-4.0 keV. For the purposes of display, these images have been smoothed by a Gaussian kernel of $\sigma=0''.25$, and $\sigma=1''.96$ panels (a) and (b), respectively.} 
\end{center}
\end{figure}

\subsection{Spectroscopy}
\subsection{Ambient Medium}
We first analyzed the outer rim of SNR DEM L71, which represents the ISM swept-up by forward shock. For this purpose, we selected eight regions and marked them as S1-S8 (Figure 2a). Shell regions contain almost 5,000 counts on average, in the 0.3-4.0 keV energy band. In order to reveal the spatial structure of X-ray emission from metal-rich ejecta gas, we selected 50 regions through seven radial directions across DEM L71. The radial regions and directions are also marked as shown in Figure 2b. Each region contains about 9,000-10,000 counts in the 0.3-4.0 keV energy band. Then, we extracted the spectra from each individual observation for all selected regions, and combined them using the CIAO script {$ combine\_spectra$}. We performed the background subtraction using the spectrum extracted from source-free regions outside of DEM L71. We binned each extracted spectrum to comprise at least 20 counts per photon energy channel. We fit each regional spectrum with a non-equilibrium ionization (NEI) plane-shock model \citep[vpshock in XSPEC;][]{Borkowski01} with two foreground absorption column components, one for the Galactic ($N_{{\rm H,Gal}}$) and the other for the LMC ($N_{{\rm H,LMC}}$). We used NEI version 2.0 in XSPEC related with ATOMDB \citep{Foster12}, which was augmented to include inner shell lines, and updated Fe-L lines \citep[see, ][]{Badenes06}. We fixed $N_{{\rm H,Gal}}$ at $1.58\times10^{21}$ cm$^{-2}$ for the direction toward DEM L71 \citep{HI4PI16} with solar abundances \citep{Anders89}. We fitted $N_{{\rm H,LMC}}$ assuming the LMC abundances \citep{Russell92, Schenck16}. We also fixed the redshift parameter at $z=8.75\times10^{-4}$ for the radial velocity (262.2 kms$^{-1}$) of the LMC \citep{McConnachie12}.

\begin{figure}
\begin{center}
\includegraphics[scale=0.50]{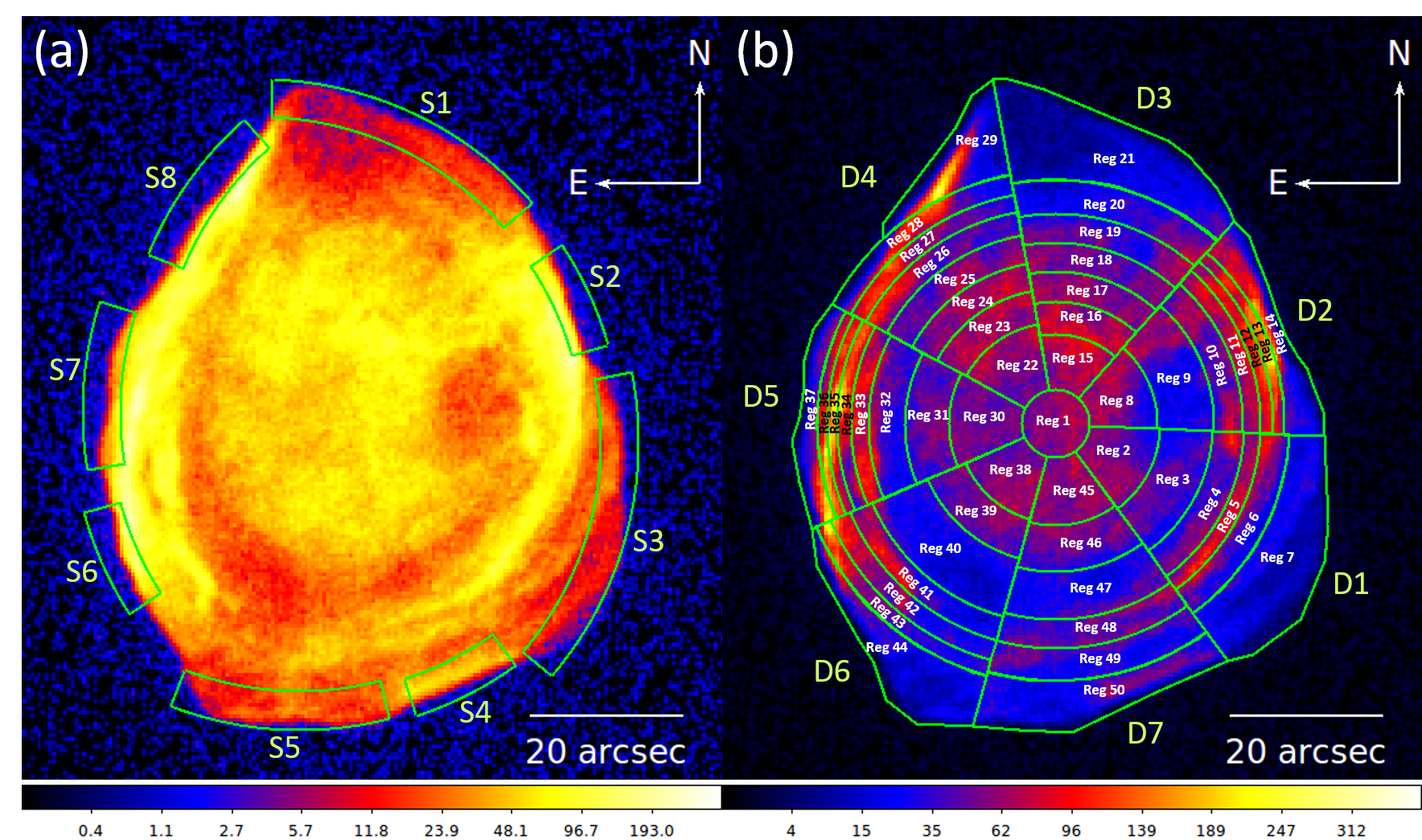}
\caption{(a) Logarithmic-scale broadband image of DEM L71 in the $0.1-4.0$ keV photon energy band. (b) The outermost eight shell regions that used to characterize the spectral nature of the swept-up ISM are marked. A square-root scale broadband image of DEM L71 in the $0.1-4.0$ keV photon energy band (b). The regions used for the spectral analysis are marked. Both images have been smoothed by a Gaussian kernel of $\sigma=0.''25$.} 
\end{center}
\end{figure}

\begin{figure}
\begin{center}
\includegraphics[scale=0.37]{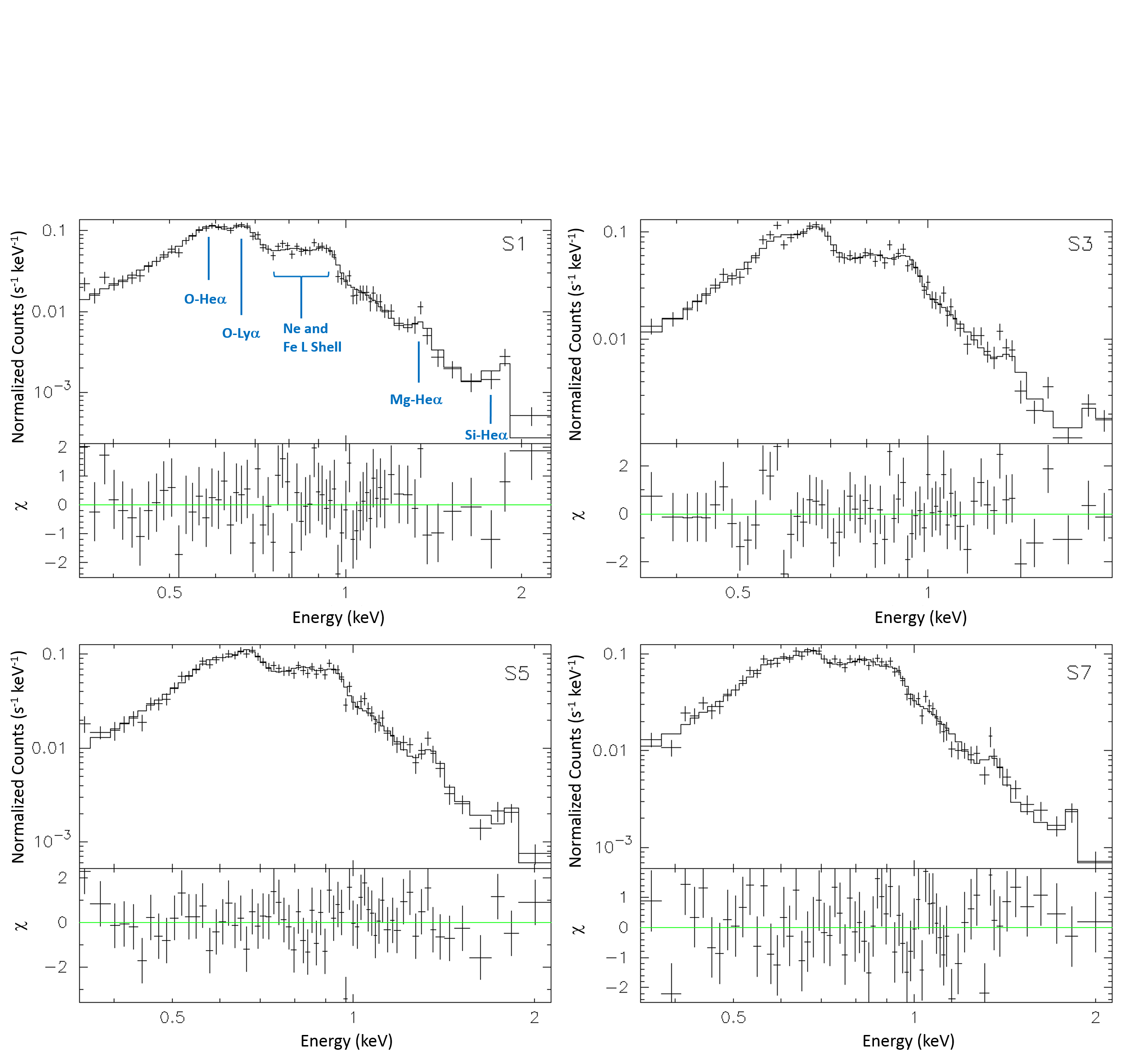}
\caption{Best-fit spectral models and residuals of the X-ray spectra from the S1, S3, S5, and S7 regions. Several atomic emission line features are marked on top left panel.} 
\end{center}
\end{figure}

We fit S1 - S8 regions spectra using an one-component plane-parallel shock (\texttt{phabs $\times$ vphabs $\times$ vpshock}) model. In Figure 3, we show the spectra of S1, S3, S5, and S7 regions with models and residuals as a sample. We initially fixed all elemental abundances at the LMC values, i.e. He = 0.89, C=0.303, S=0.31, N=0.123, Ar=0.537, Ca=0.339, Ni=0.618 \citep{Russell92}, and O=0.13, Ne=0.20, Mg=0.20, Si=0.28, Fe=0.15 \citep{Schenck16}, in the plane-shock model. Hereafter, abundances are with respect to solar values \citep{Anders89}. We varied electron temperature ($kT$, where $k$ is the Boltzmann constant), ionization timescale ($n_{\rm e}t$, where $n_{\rm e}$ is the post-shock electron density, and $t$ is the time since the gas has been shocked) and the $N_{\rm H,LMC}$ absorbing column in the LMC. The normalization parameter (a scaled volume emission measure, $EM=n_{\rm e}n_{\rm H}V$, where $n_{\rm H}$ is the postshock H density, and $V$ is the emission volume) is also varied. The reduced chi-square ($\chi^2_{\nu}$) values between $1.3-1.7$ for the model fits. Then to improve the fits we varied O, Ne, Mg, Si, Fe abundances and obtained the best-fit models for the shell regions ($\chi^2_{\nu}=0.84-1.29$). While the fitted Si abundance is consistent (within statistical uncertainties) with values given by \citet{Russell92}, the fitted abundances for O, Ne, Mg, Fe are significantly lower by a factor of $\sim  2-3$ than \citet{Russell92} values. Besides this all fitted elemental abundances consistent with \citet{Schenck16} values within statistical uncertainties. We found that the outer rim spectrum of DEM L71 is dominated by emission from the shocked low-abundant LMC ISM rather than that from the shocked metal-rich ejecta gas. The best-fit model parameters of the shell regions and their median values are listed in Table 2.

\begin{table*}
\scriptsize
  \centering
  \caption{Summary of spectral model fits to the eight shell regions of DEM L71. The median shell values are given in the last line.}
    \begin{tabular}{ccccccccccc}
\hline
Region    &    $n_{\rm H}$             &   $kT$  & $n_{\rm e}t$                & $EM$                 & O    & Ne  & Mg  & Si & Fe & $\chi^{2}_{\nu}$\\
    &($10^{21}$cm$^{-2}$)  &  (keV)  & ($10^{11}$cm$^{-3}$s) & ($10^{57}$cm$^{-3}$) &      &     &     &    &    &                 \\
\hline
S1 & $0.20_{-0.10}^{+0.10}$ & $0.31_{-0.05}^{+0.05}$ & $ 2.79_{-1.41}^{+2.48}$ & $ 27.16_{-9.16}^{+11.97}$ & $0.11_{-0.02}^{+0.02}$ & $0.18_{-0.05}^{+0.02}$ & $0.12_{-0.07}^{+0.08}$ & $0.52_{-0.23}^{+0.46}$ & $0.08_{-0.02}^{+0.03}$ & 0.96 \\
S2 & $0.30_{-0.10}^{+0.20}$ & $0.26_{-0.04}^{+0.04}$ & $ 4.86_{-2.54}^{+4.62}$ & $ 48.72_{-18.51}^{+24.27}$ & $0.09_{-0.01}^{+0.20}$ & $0.16_{-0.03}^{+0.04}$ & $0.14_{-0.07}^{+0.09}$ & $0.67_{-0.33}^{+0.31}$ & $0.13_{-0.03}^{+0.04}$ & 1.28 \\
S3 & $0.20_{-0.10}^{+0.20}$ & $0.34_{-0.08}^{+0.04}$ & $ 3.29_{-1.31}^{+21.93}$ & $ 23.24_{-5.82}^{+23.91}$ & $0.12_{-0.02}^{+0.20}$ & $0.17_{-0.04}^{+0.03}$ & $0.09_{-0.06}^{+0.08}$ & $0.33_{-0.10}^{+0.34}$ & $0.07_{-0.02}^{+0.03}$ & 1.29 \\
S4 & $0.10_{-0.10}^{+0.30}$ & $0.43_{-0.07}^{+0.08}$ & $ 2.10_{-1.02}^{+2.91}$ & $ 12.33_{-4.03}^{+8.35}$ & $0.13_{-0.03}^{+0.30}$ & $0.25_{-0.03}^{+0.03}$ & $0.19_{-0.07}^{+0.08}$ & $0.41_{-0.13}^{+0.30}$ & $0.12_{-0.03}^{+0.05}$ & 1.08 \\
S5 & $0.10_{-0.10}^{+0.10}$ & $0.36_{-0.06}^{+0.07}$ & $ 2.50_{-1.23}^{+6.69}$ & $ 18.52_{-6.25}^{+13.98}$ & $0.11_{-0.02}^{+0.20}$ & $0.18_{-0.03}^{+0.04}$ & $0.14_{-0.06}^{+0.07}$ & $0.26_{-0.13}^{+0.18}$ & $0.10_{-0.02}^{+0.03}$ & 1.12 \\
S6 & $0.30_{-0.20}^{+0.20}$ & $0.36_{-0.06}^{+0.05}$ & $ 2.12_{-0.89}^{+2.63}$ & $ 24.74_{-7.96}^{+19.71}$ & $0.10_{-0.02}^{+0.20}$ & $0.14_{-0.04}^{+0.03}$ & $0.12_{-0.05}^{+0.06}$ & $0.27_{-0.15}^{+0.19}$ & $0.11_{-0.03}^{+0.02}$ & 0.97 \\
S7 & $0.20_{-0.16}^{+0.30}$ & $0.36_{-0.07}^{+0.05}$ & $ 2.44_{-1.05}^{+4.12}$ & $ 18.14_{-5.26}^{+20.29}$ & $0.11_{-0.03}^{+0.02}$ & $0.20_{-0.05}^{+0.04}$ & $0.12_{-0.06}^{+0.06}$ & $0.33_{-0.18}^{+0.25}$ & $0.16_{-0.04}^{+0.05}$ & 1.16 \\
S8 & $0.20_{-0.15}^{+0.20}$ & $0.30_{-0.04}^{+0.04}$ & $ 2.20_{-1.02}^{+1.58}$ & $ 28.18_{-8.37}^{+17.39}$ & $0.10_{-0.02}^{+0.10}$ & $0.15_{-0.03}^{+0.03}$ & $0.16_{-0.07}^{+0.08}$ & $0.36_{-0.15}^{+0.34}$ & $0.18_{-0.04}^{+0.05}$ & 0.84 \\
\hline
Median & $0.20_{-0.20}^{+0.10}$ & $0.35_{-0.03}^{+0.04}$ & $ 2.47_{-1.48}^{+2.01}$ & $ 23.99_{-7.11}^{+18.55}$ & $0.11_{-0.01}^{+0.01}$ & $0.18_{-0.02}^{+0.02}$ & $0.14_{-0.04}^{+0.05}$ & $0.34_{-0.11}^{+0.13}$ & $0.12_{-0.01}^{+0.01}$ & ---  \\
\hline
    \end{tabular}%
\\
Note: Abundances are with respect to solar \citep{Anders89}. Uncertainties are at the 90\% confidence level. The Galactic column $N_{{\rm H,Gal}}$ is fixed at $1.74\times10^{21}$ cm$^{-2}$ \citep{HI4PI16}. 
\end{table*}%

\begin{figure}
\begin{center}
\includegraphics[scale=0.40]{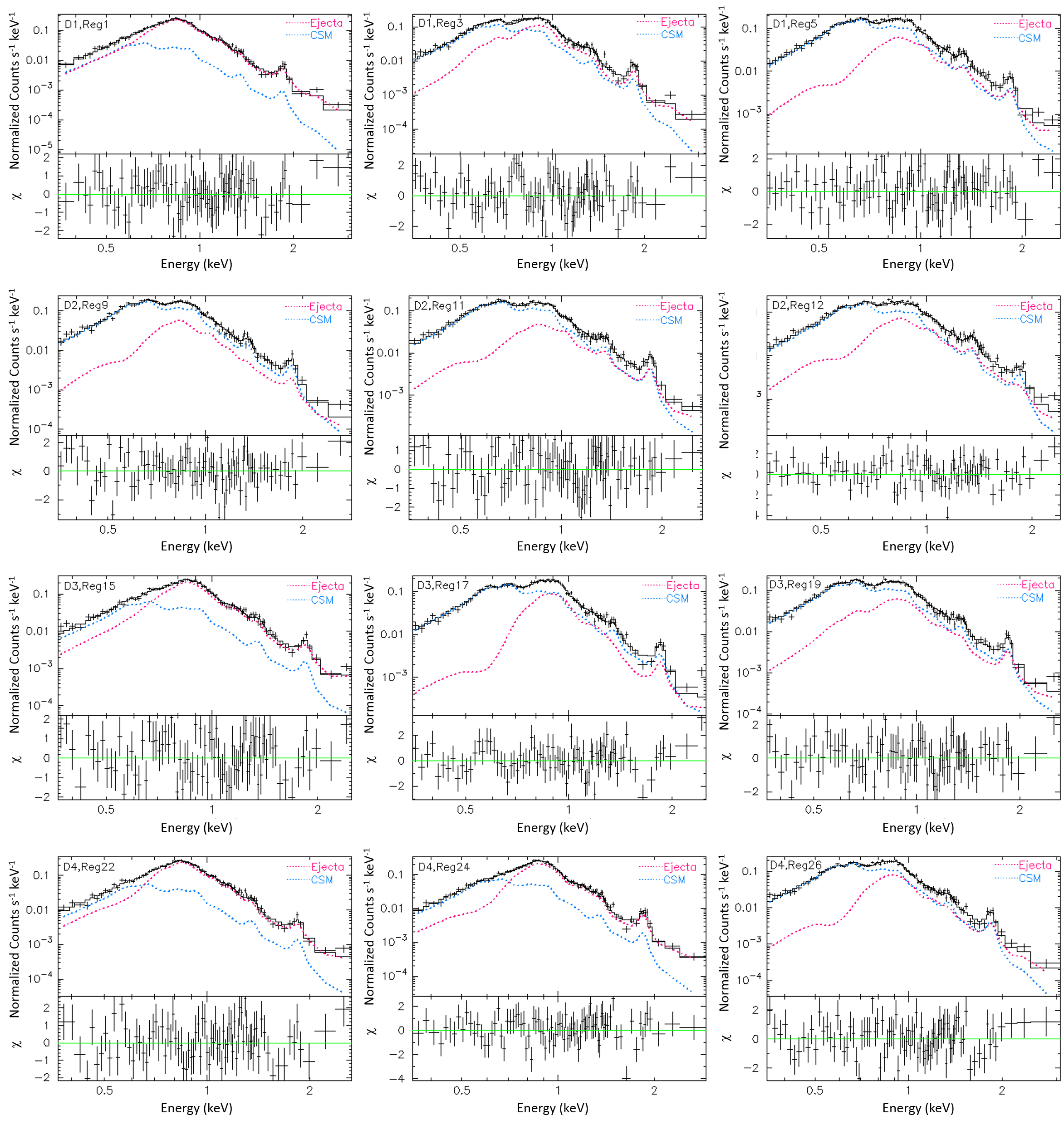}
\caption{
 A set of sample best-fit models and residuals of X-ray spectra from selected regions shown in Figure 2b.}
\end{center}
\end{figure}
\begin{figure}
\begin{center}
\includegraphics[scale=0.40]{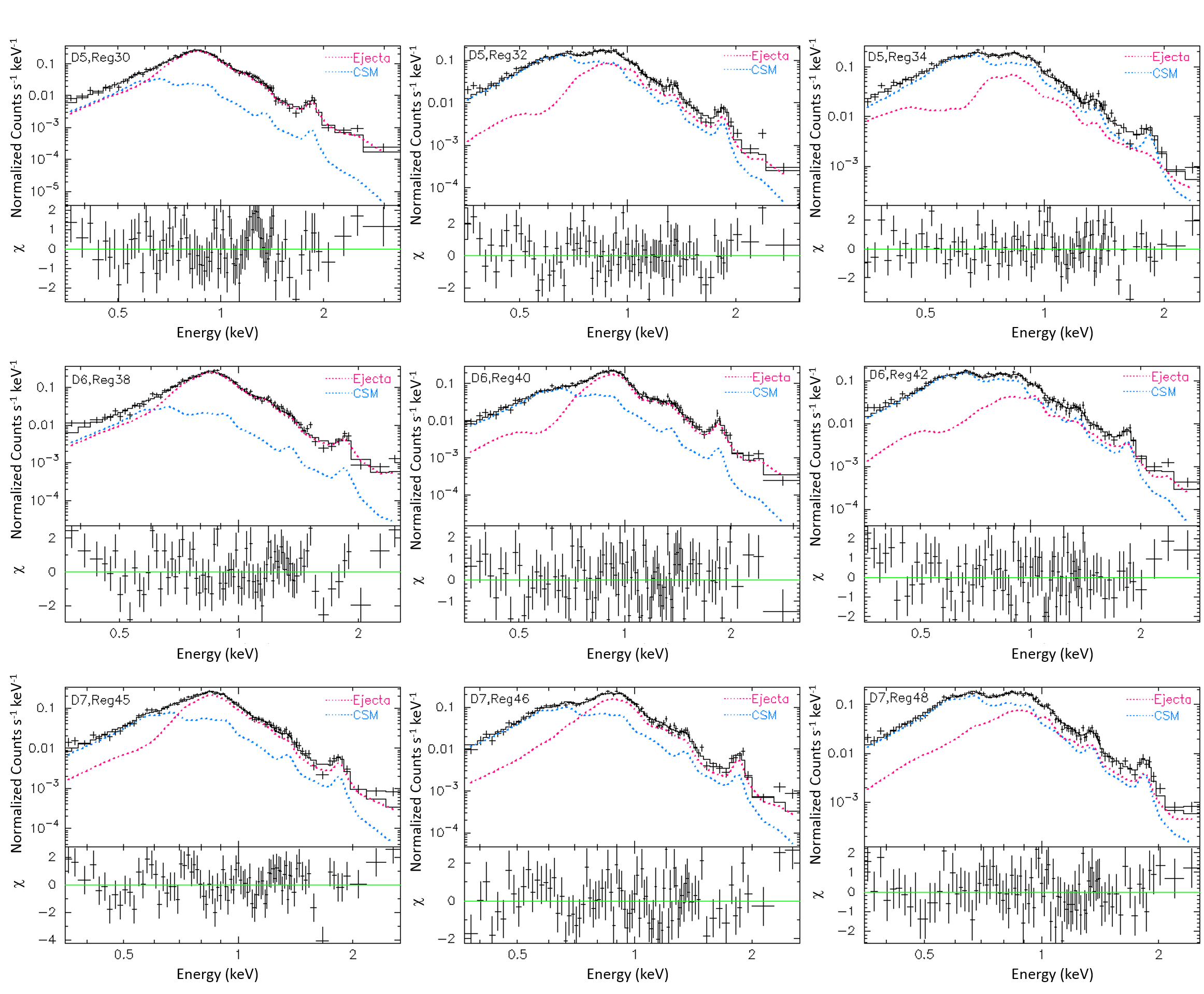}
\caption{
 A set of sample best-fit models and residuals of X-ray spectra from selected regions shown in Figure 2b.}
\end{center}
\end{figure}

\begin{figure*}
\begin{center}
\includegraphics[scale=0.80]{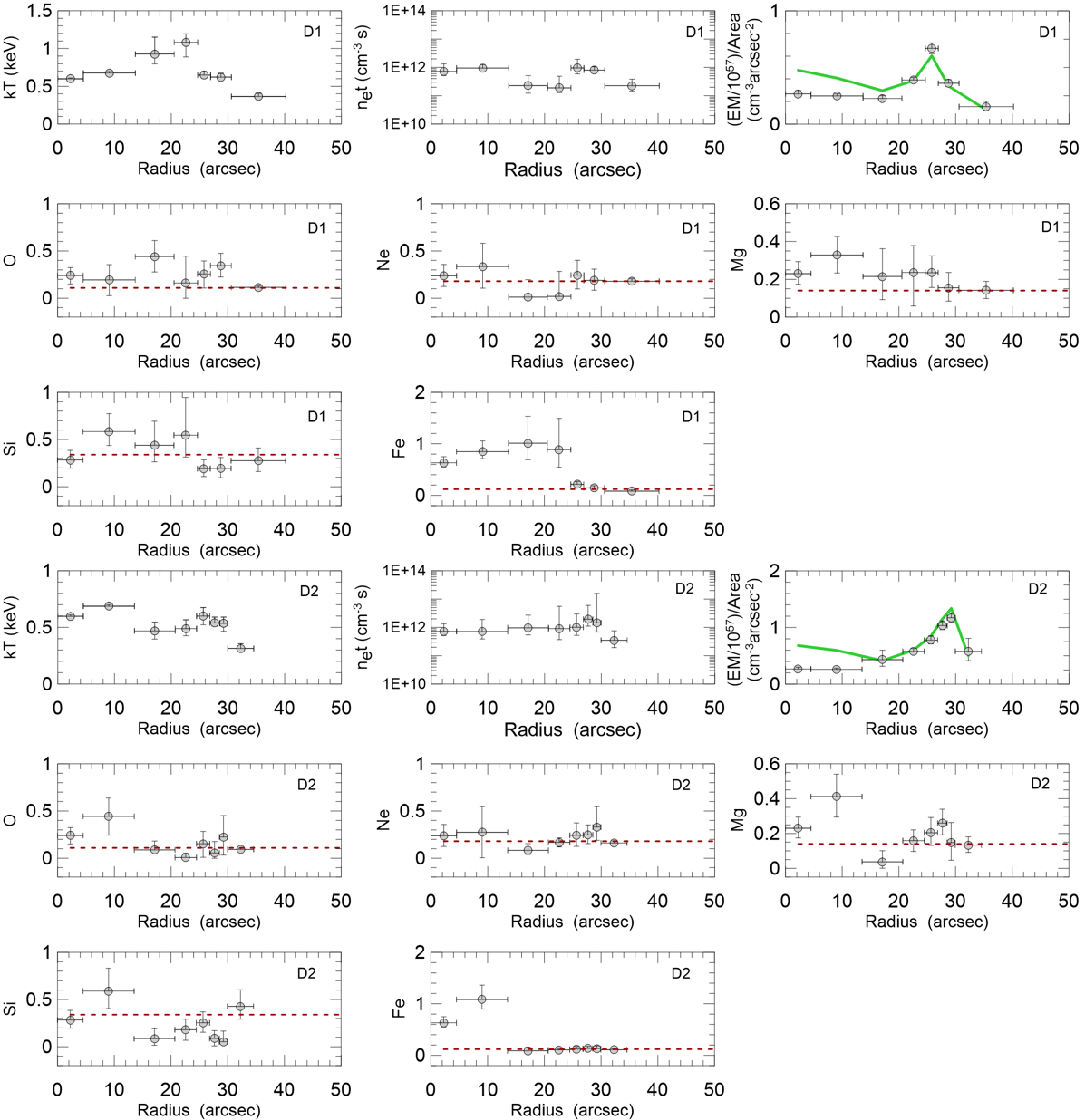}
\caption{Best-fit spectral parameters with error bars along the radius of DEM L71. The parameters for ejecta component are electron temperature ($kT$), ionization timescale ($n_et$), emission measure ($EM$), and O, Ne, Mg, Si, and Fe abundances in the D1 and D2 directions of DEM L71. In $EM$ plots the broadband surface brightness profile is overlaid with a green line. The red dashed lines in the abundance panels are the mean shell abundances for each element. Abundances are with respect to solar \citep{Anders89}.}
\end{center}
\end{figure*}

\begin{figure*}
\begin{center}
\includegraphics[scale=0.80]{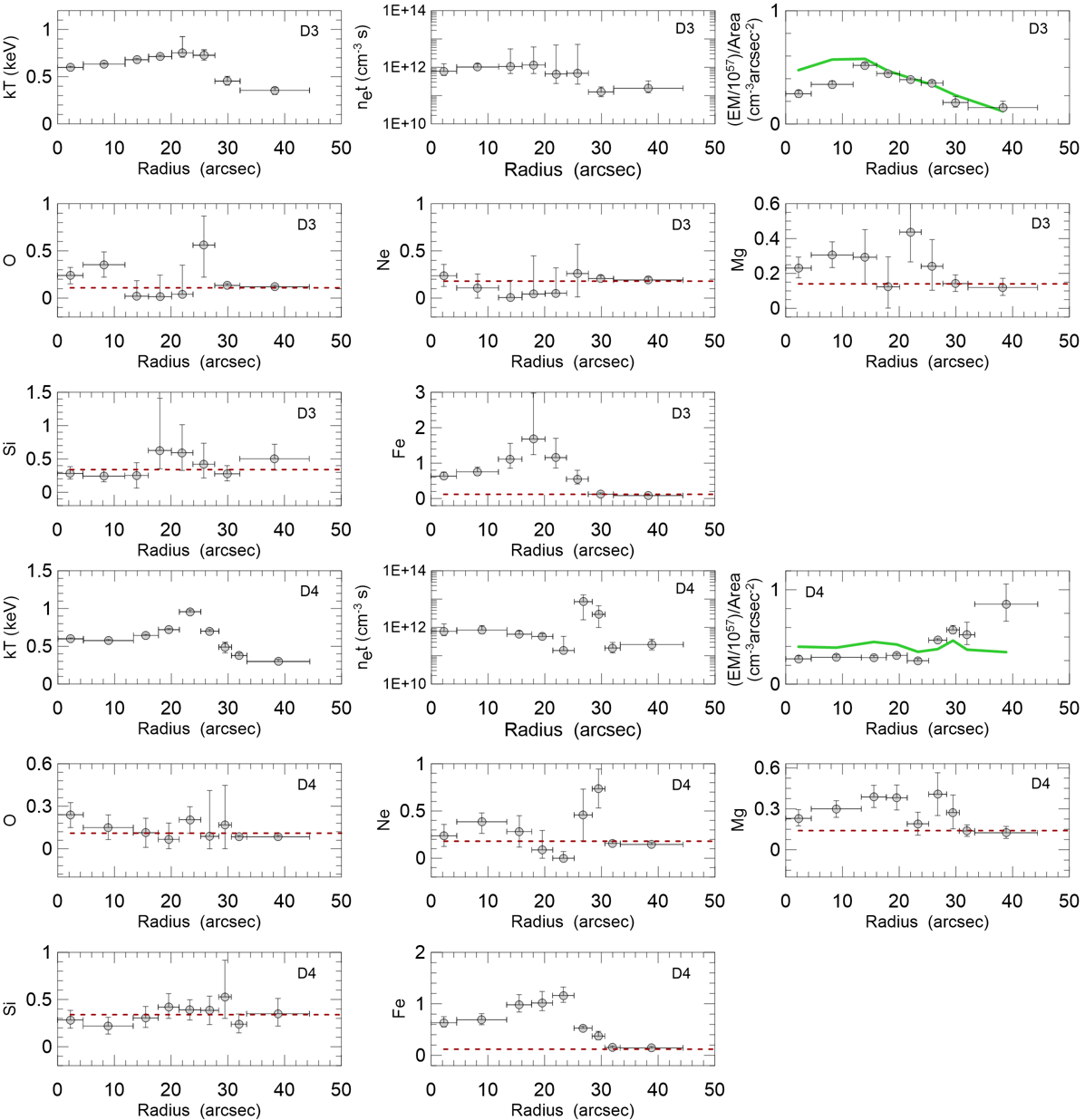}
\caption{Same as spectral parameters in Figure 6 but for the D3 and D4 directions of DEM L71.}
\end{center}
\end{figure*}

\begin{figure*}
\begin{center}
\includegraphics[scale=0.80]{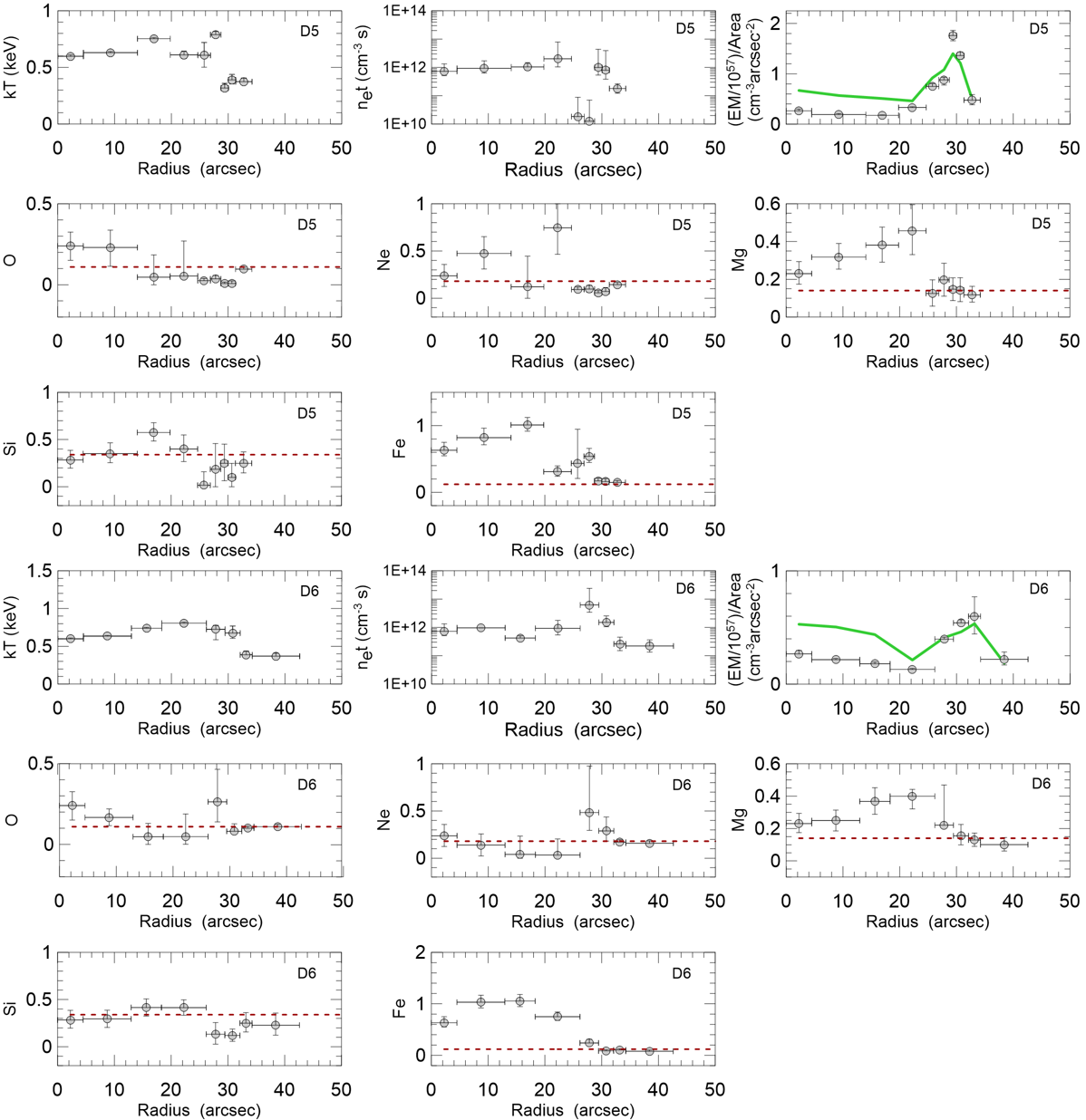}
\caption{Same as spectral parameters in Figure 6 but for the D5 and D6 directions of DEM L71.}
\end{center}
\end{figure*}

\begin{figure*}
\begin{center}
\includegraphics[scale=0.80]{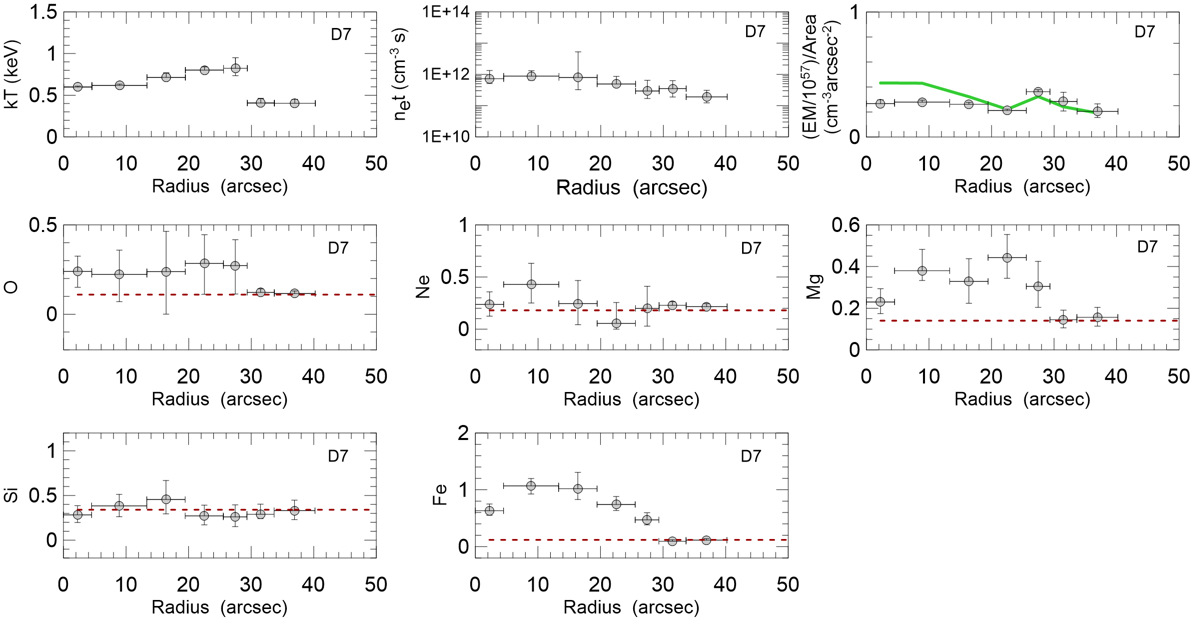}
\caption{Same as spectral parameters in Figure 6 but for the D7 direction of DEM L71.}
\end{center}
\end{figure*}

\newpage
\subsection{Metal-Rich Ejecta}

We perform an extensive spatially resolved spectral analysis of its X-ray emission based on 50 radial and azimuthal regional (see Figure 2b) spectra to study the detailed spatial distribution of metal-rich ejecta in DEM L71. The outermost regions of the remnant can be modeled with a single component plane shock model, and the spectral parameters for these regions are compatible with median shell values. The spectra of more inside regions cannot be fit by a single shock model with abundances fixed at the values that we estimated for the mean shell (i.e. $\chi^2_{\nu}>2.0$). These show the existence of an additional shock component, likely representing the emission from the shocked metal-rich ejecta gas, superposed with the projected shell emission. Therefore, we performed a two-component NEI shock model (\texttt{phabs $\times$ vphabs $\times$ (vpshock+vpshock)}) to fit these spectra, one for the underlying mean shell spectrum and the other responsible for the metal-rich ejecta component. We fixed $N_{{\rm H,LMC}}$ at the mean shell value. We also fixed all model parameters, except for normalization, of the underlying swept-up ISM component at the values for the mean shell (Table 2). For the second shock component we first varied $kT$, $n_{\rm e}t$, and normalization and fixed the elemental abundances at the mean shell values. The fit for each regional spectrum was not statistically acceptable ($\chi^2_{\nu}>2.0$) because the model was not able to reproduce emission lines from various elements. Then we thawed elemental abundances for the second component to improve the each regional spectral fit. With abundances for O, Ne, Mg, Si, and Fe varied, our spectral model fits significantly improved ($\chi^2_{\nu}<1.6$). In Figures 4-5, we show some example spectra extracted from the regions marked in Figure 2b with best-fit models and residuals. The Fe abundance in the central parts of the remnant is significantly enhanced compared to those of the mean shell values. O, Ne, and Mg abundances are generally consistent with mean shell values within statistical uncertainties. The best-fit model parameters for radial regions are listed in Table 3, and radial profiles of spectral parameters are shown in Figures 6-9.

\begin{table*}
\scriptsize
  \centering
  \caption{Spectral parameters from NEI shock model fits for 50 radial and azimuthal regions spectra of DEM L71.}
    \begin{tabular}{cccccccccccc}
\hline
Direction &    Region & $r$     & $kT$   & $n_{\rm e}t$   & $EM$        & O                     & Ne   & Mg  & Si & Fe & $\chi^{2}_{\nu}$\\
          &    & ($''$)  & (keV)  & ($10^{11}$cm$^{-3}$s) & ($10^{57}$cm$^{-3}$) &      &     &    &    & & \\
\hline
			D1 & 1      & 2.25     & $0.60_{-0.01}^{+0.01}$ & $7.27_{-2.20}^{+6.13}$     & $9.33_{-1.42}^{+1.45}$    & $0.24_{-0.09}^{+0.09}$ & $0.24_{-0.11}^{+0.12}$ & $0.23_{-0.06}^{+0.06}$ & $0.28_{-0.09}^{+0.10}$ & $0.63_{-0.08}^{+0.12}$ & 1.03    \\
		& 2      & 9.09     & $0.68_{-0.01}^{+0.02}$ & $9.50_{-1.95}^{+3.20}$     & $5.75_{-1.10}^{+1.04}$    & $0.19_{-0.17}^{+0.16}$ & $0.34_{-0.23}^{+0.25}$ & $0.33_{-0.10}^{+0.10}$ & $0.58_{-0.15}^{+0.19}$ & $0.85_{-0.14}^{+0.20}$ & 1.23    \\
		& 3      & 17.10     & $0.93_{-0.13}^{+0.23}$ & $2.28_{-1.06}^{+2.79}$     & $2.79_{-1.14}^{+1.49}$    & $0.44_{-0.16}^{+0.17}$ & $0.01_{-0.01}^{+0.18}$ & $0.22_{-0.12}^{+0.15}$ & $0.44_{-0.18}^{+0.26}$ & $1.01_{-0.32}^{+0.52}$ & 1.24    \\
		& 4      & 22.59    & $1.08_{-0.19}^{+0.12}$ & $1.88_{-0.60}^{+3.01}$     & $1.83_{-0.78}^{+1.40}$    & $0.16_{-0.16}^{+0.28}$ & $0.02_{-0.01}^{+0.26}$ & $0.24_{-0.18}^{+0.14}$ & $0.55_{-0.23}^{+0.40}$ & $0.89_{-0.34}^{+0.61}$ & 1.32    \\
		& 5      & 25.79   & $0.65_{-0.04}^{+0.05}$ & $9.53_{-3.65}^{+9.92}$     & $8.23_{-1.71}^{+1.88}$    & $0.26_{-0.15}^{+0.14}$ & $0.24_{-0.15}^{+0.16}$ & $0.23_{-0.08}^{+0.09}$ & $0.19_{-0.08}^{+0.10}$ & $0.21_{-0.04}^{+0.05}$ & 1.18    \\
		& 6      & 28.76   & $0.62_{-0.04}^{+0.05}$ & $8.03_{-1.61}^{+2.43}$     & $7.73_{-1.87}^{+1.95}$    & $0.34_{-0.12}^{+0.13}$ & $0.19_{-0.10}^{+0.12}$ & $0.15_{-0.07}^{+0.08}$ & $0.19_{-0.10}^{+0.12}$ & $0.14_{-0.03}^{+0.04}$ & 1.54    \\
		& 7      & 35.40     & $0.36_{-0.04}^{+0.04}$ & $2.24_{-0.80}^{+1.59}$     & $38.16_{-8.87}^{+11.56}$  & $0.11_{-0.01}^{+0.01}$ & $0.18_{-0.02}^{+0.03}$ & $0.14_{-0.04}^{+0.05}$ & $0.28_{-0.11}^{+0.13}$ & $0.08_{-0.02}^{+0.02}$ & 1.36    \\ \hline
		D2 & 8      & 9.00        & $0.69_{-0.01}^{+0.01}$ & $7.08_{-3.19}^{+11.84}$    & $4.09_{-0.83}^{+0.86}$    & $0.44_{-0.20}^{+0.19}$ & $0.27_{-0.27}^{+0.27}$ & $0.41_{-0.12}^{+0.13}$ & $0.59_{-0.19}^{+0.24}$ & $1.08_{-0.18}^{+0.28}$ & 1.29    \\
		& 9      & 17.10     & $0.47_{-0.07}^{+0.08}$ & $9.69_{-4.28}^{+17.36}$    & $21.40_{-11.03}^{+18.51}$ & $0.09_{-0.04}^{+0.09}$ & $0.08_{-0.04}^{+0.07}$ & $0.04_{-0.04}^{+0.06}$ & $0.08_{-0.07}^{+0.11}$ & $0.09_{-0.03}^{+0.07}$ & 1.24    \\
		& 10     & 22.59    & $0.49_{-0.06}^{+0.08}$ & $9.07_{-5.38}^{+46.45}$    & $10.86_{-2.77}^{+3.61}$   & $0.01_{-0.01}^{+0.05}$ & $0.17_{-0.05}^{+0.05}$ & $0.16_{-0.06}^{+0.06}$ & $0.18_{-0.11}^{+0.11}$ & $0.11_{-0.02}^{+0.02}$ & 1.37    \\
		& 11     & 25.65    & $0.60_{-0.08}^{+0.08}$ & $10.06_{-4.78}^{+19.64}$   & $9.75_{-2.63}^{+3.66}$    & $0.15_{-0.14}^{+0.13}$ & $0.24_{-0.12}^{+0.13}$ & $0.20_{-0.07}^{+0.09}$ & $0.25_{-0.10}^{+0.12}$ & $0.12_{-0.03}^{+0.05}$ & 0.98    \\
		& 12     & 27.67   & $0.54_{-0.03}^{+0.05}$ & $19.33_{-8.22}^{+40.73}$   & $12.80_{-2.72}^{+3.21}$   & $0.06_{-0.06}^{+0.12}$ & $0.25_{-0.09}^{+0.11}$ & $0.26_{-0.07}^{+0.08}$ & $0.09_{-0.08}^{+0.08}$ & $0.14_{-0.03}^{+0.04}$ & 1.54    \\
		& 13     & 29.23   & $0.53_{-0.07}^{+0.06}$ & $14.54_{-7.65}^{+144.01}$  & $7.24_{-2.02}^{+2.51}$    & $0.22_{-0.19}^{+0.23}$ & $0.33_{-0.14}^{+0.22}$ & $0.15_{-0.10}^{+0.11}$ & $0.05_{-0.02}^{+0.11}$ & $0.13_{-0.04}^{+0.06}$ & 1.54    \\
		& 14     & 32.26   & $0.31_{-0.03}^{+0.04}$ & $3.52_{-1.59}^{+4.07}$     & $63.59_{-18.28}^{+24.85}$ & $0.09_{-0.01}^{+0.01}$ & $0.16_{-0.02}^{+0.02}$ & $0.13_{-0.04}^{+0.05}$ & $0.43_{-0.13}^{+0.18}$ & $0.11_{-0.02}^{+0.02}$ & 1.32    \\ \hline
		D3 & 15     & 8.19     & $0.63_{-0.01}^{+0.01}$ & $10.24_{-1.94}^{+3.03}$    & $7.39_{-1.15}^{+1.18}$    & $0.35_{-0.13}^{+0.13}$ & $0.11_{-0.11}^{+0.15}$ & $0.31_{-0.07}^{+0.07}$ & $0.24_{-0.08}^{+0.11}$ & $0.75_{-0.10}^{+0.13}$ & 1.19    \\
		& 16     & 13.95    & $0.68_{-0.02}^{+0.01}$ & $10.73_{-4.71}^{+33.60}$   & $2.78_{-0.76}^{+0.76}$    & $0.02_{-0.01}^{+0.16}$ & $0.01_{-0.01}^{+0.18}$ & $0.29_{-0.16}^{+0.16}$ & $0.25_{-0.19}^{+0.19}$ & $1.11_{-0.26}^{+0.44}$ & 1.11    \\
		& 17     & 18.05   & $0.71_{-0.03}^{+0.01}$ & $12.10_{-6.05}^{+39.73}$   & $2.30_{-0.18}^{+0.11}$    & $0.02_{-0.01}^{+0.23}$ & $0.04_{-0.01}^{+0.41}$ & $0.12_{-0.12}^{+0.17}$ & $0.63_{-0.27}^{+0.78}$ & $1.68_{-0.44}^{+1.30}$ & 1.12    \\
		& 18     & 21.98  & $0.75_{-0.05}^{+0.17}$ & $5.79_{-3.10}^{+54.38}$    & $2.13_{-0.65}^{+0.65}$    & $0.04_{-0.01}^{+0.31}$ & $0.05_{-0.02}^{+0.27}$ & $0.44_{-0.17}^{+0.19}$ & $0.59_{-0.26}^{+0.42}$ & $1.16_{-0.29}^{+0.54}$ & 1.54    \\
		& 19     & 25.79  & $0.73_{-0.05}^{+0.06}$ & $6.15_{-3.61}^{+57.83}$    & $2.79_{-0.85}^{+0.85}$    & $0.56_{-0.34}^{+0.31}$ & $0.26_{-0.25}^{+0.31}$ & $0.24_{-0.14}^{+0.15}$ & $0.42_{-0.21}^{+0.32}$ & $0.55_{-0.14}^{+0.25}$ & 0.95    \\
		& 20     & 29.91  & $0.45_{-0.04}^{+0.05}$ & $1.34_{-0.43}^{+0.67}$     & $22.80_{-4.92}^{+6.39}$   & $0.13_{-0.02}^{+0.02}$ & $0.21_{-0.03}^{+0.03}$ & $0.14_{-0.04}^{+0.05}$ & $0.28_{-0.11}^{+0.12}$ & $0.12_{-0.02}^{+0.03}$ & 1.47    \\
		& 21     & 38.26   & $0.36_{-0.04}^{+0.03}$ & $1.82_{-0.57}^{+1.50}$     & $35.69_{-7.11}^{+14.02}$  & $0.12_{-0.01}^{+0.01}$ & $0.19_{-0.03}^{+0.03}$ & $0.12_{-0.05}^{+0.05}$ & $0.50_{-0.17}^{+0.22}$ & $0.08_{-0.02}^{+0.02}$ & 1.47    \\ \hline
		D4 & 22     & 8.91     & $0.58_{-0.01}^{+0.01}$ & $8.06_{-1.96}^{+3.53}$     & $9.00_{-0.68}^{+1.54}$    & $0.15_{-0.09}^{+0.09}$ & $0.39_{-0.12}^{+0.09}$ & $0.30_{-0.06}^{+0.06}$ & $0.22_{-0.08}^{+0.09}$ & $0.69_{-0.10}^{+0.12}$ & 1.16    \\
		& 23     & 15.52   & $0.64_{-0.01}^{+0.01}$ & $5.77_{-1.11}^{+1.77}$     & $6.56_{-1.10}^{+1.05}$    & $0.11_{-0.10}^{+0.10}$ & $0.28_{-0.16}^{+0.17}$ & $0.39_{-0.08}^{+0.09}$ & $0.31_{-0.10}^{+0.12}$ & $0.98_{-0.14}^{+0.20}$ & 1.41    \\
		& 24     & 19.57   & $0.72_{-0.01}^{+0.01}$ & $4.76_{-0.84}^{+1.24}$     & $5.50_{-0.97}^{+0.94}$    & $0.07_{-0.07}^{+0.12}$ & $0.09_{-0.09}^{+0.20}$ & $0.38_{-0.09}^{+0.09}$ & $0.42_{-0.12}^{+0.14}$ & $1.02_{-0.15}^{+0.22}$ & 1.22    \\
		& 25     & 23.31    & $0.96_{-0.03}^{+0.02}$ & $1.53_{-0.35}^{+3.31}$     & $3.12_{-0.34}^{+0.34}$    & $0.21_{-0.10}^{+0.09}$ & $0.00_{-0.00}^{+0.07}$ & $0.19_{-0.08}^{+0.08}$ & $0.39_{-0.11}^{+0.11}$ & $1.16_{-0.13}^{+0.16}$ & 1.28    \\
		& 26     & 26.78   & $0.70_{-0.03}^{+0.02}$ & $80.21_{-61.65}^{+60.21}$  & $4.00_{-0.29}^{+0.33}$    & $0.09_{-0.09}^{+0.32}$ & $0.46_{-0.27}^{+0.27}$ & $0.41_{-0.16}^{+0.16}$ & $0.39_{-0.15}^{+0.15}$ & $0.53_{-0.03}^{+0.03}$ & 1.18    \\
		& 27     & 29.48   & $0.49_{-0.08}^{+0.06}$ & $28.79_{-18.85}^{+28.79}$  & $6.40_{-1.51}^{+2.44}$    & $0.17_{-0.17}^{+0.28}$ & $0.74_{-0.20}^{+0.21}$ & $0.27_{-0.12}^{+0.13}$ & $0.53_{-0.23}^{+0.39}$ & $0.37_{-0.07}^{+0.09}$ & 1.43    \\
		& 28     & 31.95    & $0.38_{-0.03}^{+0.03}$ & $1.86_{-0.62}^{+1.07}$     & $40.09_{-8.09}^{+10.14}$  & $0.08_{-0.01}^{+0.01}$ & $0.16_{-0.02}^{+0.02}$ & $0.14_{-0.04}^{+0.04}$ & $0.24_{-0.09}^{+0.11}$ & $0.15_{-0.03}^{+0.03}$ & 1.38    \\
		& 29     & 38.85    & $0.30_{-0.02}^{+0.03}$ & $2.49_{-0.86}^{+1.35}$     & $72.56_{-15.37}^{+18.29}$ & $0.08_{-0.01}^{+0.01}$ & $0.15_{-0.02}^{+0.02}$ & $0.12_{-0.04}^{+0.05}$ & $0.35_{-0.13}^{+0.16}$ & $0.15_{-0.02}^{+0.03}$ & 1.27    \\ \hline
		D5 & 30     & 9.27     & $0.63_{-0.01}^{+0.01}$ & $9.27_{-2.98}^{+7.69}$     & $7.95_{-1.18}^{+1.11}$    & $0.23_{-0.12}^{+0.11}$ & $0.47_{-0.16}^{+0.18}$ & $0.32_{-0.06}^{+0.07}$ & $0.35_{-0.10}^{+0.11}$ & $0.82_{-0.11}^{+0.14}$ & 1.17    \\
		& 31     & 16.92    & $0.75_{-0.02}^{+0.01}$ & $10.36_{-2.43}^{+4.32}$    & $6.67_{-0.15}^{+0.20}$    & $0.05_{-0.05}^{+0.14}$ & $0.12_{-0.12}^{+0.32}$ & $0.38_{-0.09}^{+0.10}$ & $0.58_{-0.09}^{+0.10}$ & $1.01_{-0.09}^{+0.11}$ & 1.03    \\
		& 32     & 22.23    & $0.61_{-0.03}^{+0.03}$ & $19.95_{-9.63}^{+59.24}$   & $6.54_{-0.69}^{+0.64}$    & $0.05_{-0.02}^{+0.22}$ & $0.75_{-0.28}^{+0.27}$ & $0.46_{-0.13}^{+0.14}$ & $0.40_{-0.13}^{+0.15}$ & $0.31_{-0.07}^{+0.09}$ & 1.37    \\
		& 33     & 25.78   & $0.61_{-0.10}^{+0.11}$ & $0.18_{-0.04}^{+0.69}$     & $10.09_{-2.05}^{+2.23}$   & $0.02_{-0.02}^{+0.01}$ & $0.09_{-0.03}^{+0.03}$ & $0.13_{-0.07}^{+0.07}$ & $0.02_{-0.01}^{+0.14}$ & $0.43_{-0.22}^{+0.51}$ & 1.49    \\
		& 34     & 27.81    & $0.79_{-0.03}^{+0.03}$ & $0.13_{-0.01}^{+0.57}$     & $6.20_{-0.91}^{+1.49}$    & $0.04_{-0.02}^{+0.02}$ & $0.10_{-0.04}^{+0.04}$ & $0.20_{-0.09}^{+0.09}$ & $0.19_{-0.19}^{+0.27}$ & $0.54_{-0.09}^{+0.12}$ & 1.40    \\
		& 35     & 29.38   & $0.32_{-0.03}^{+0.04}$ & $9.89_{-4.57}^{+33.72}$    & $29.60_{-3.34}^{+3.34}$   & $0.01_{-0.01}^{+0.01}$ & $0.06_{-0.02}^{+0.02}$ & $0.15_{-0.06}^{+0.06}$ & $0.25_{-0.19}^{+0.20}$ & $0.17_{-0.03}^{+0.04}$ & 1.53    \\
		& 36     & 30.67   & $0.39_{-0.04}^{+0.05}$ & $7.93_{-4.12}^{+30.79}$    & $17.15_{-2.07}^{+2.20}$   & $0.01_{-0.01}^{+0.02}$ & $0.07_{-0.04}^{+0.04}$ & $0.14_{-0.06}^{+0.07}$ & $0.10_{-0.10}^{+0.15}$ & $0.16_{-0.03}^{+0.04}$ & 1.43    \\
		& 37     & 32.76    & $0.37_{-0.03}^{+0.03}$ & $1.79_{-0.56}^{+0.84}$     & $36.76_{-7.14}^{+8.41}$   & $0.10_{-0.01}^{+0.01}$ & $0.14_{-0.02}^{+0.02}$ & $0.12_{-0.04}^{+0.04}$ & $0.25_{-0.10}^{+0.12}$ & $0.15_{-0.02}^{+0.03}$ & 1.14    \\ \hline
		D6 & 38     & 8.73     & $0.64_{-0.01}^{+0.01}$ & $9.67_{-2.01}^{+3.41}$     & $6.50_{-0.21}^{+0.21}$    & $0.17_{-0.05}^{+0.05}$ & $0.14_{-0.11}^{+0.12}$ & $0.25_{-0.06}^{+0.06}$ & $0.30_{-0.09}^{+0.09}$ & $1.03_{-0.11}^{+0.14}$ & 1.48    \\
		& 39     & 15.62   & $0.74_{-0.02}^{+0.01}$ & $4.24_{-0.61}^{+0.82}$     & $5.86_{-0.12}^{+0.12}$    & $0.05_{-0.05}^{+0.08}$ & $0.04_{-0.04}^{+0.19}$ & $0.37_{-0.08}^{+0.08}$ & $0.42_{-0.09}^{+0.09}$ & $1.05_{-0.11}^{+0.13}$ & 1.31    \\
		& 40     & 22.23    & $0.81_{-0.01}^{+0.01}$ & $9.26_{-3.85}^{+8.41}$     & $6.70_{-0.29}^{+0.24}$    & $0.05_{-0.05}^{+0.14}$ & $0.03_{-0.01}^{+0.17}$ & $0.40_{-0.08}^{+0.04}$ & $0.41_{-0.08}^{+0.08}$ & $0.75_{-0.07}^{+0.09}$ & 1.18    \\
	& 41     & 27.82  & $0.72_{-0.14}^{+0.05}$ & $61.81_{-27.18}^{+178.63}$ & $5.09_{-0.78}^{+0.78}$  & $0.26_{-0.12}^{+0.20}$ & $0.48_{-0.19}^{+0.49}$ & $0.22_{-0.02}^{+0.25}$ & $0.13_{-0.11}^{+0.12}$ & $0.24_{-0.05}^{+0.07}$ & 1.46 \\
		& 42     & 30.79  & $0.67_{-0.07}^{+0.09}$ & $15.01_{-4.50}^{+10.33}$   & $10.50_{-1.51}^{+1.66}$ & $0.08_{-0.01}^{+0.05}$ & $0.29_{-0.11}^{+0.15}$ & $0.16_{-0.06}^{+0.07}$ & $0.12_{-0.06}^{+0.07}$ & $0.08_{-0.02}^{+0.02}$ & 1.33 \\
		& 43     & 33.18  & $0.39_{-0.03}^{+0.05}$ & $2.60_{-1.09}^{+1.93}$     & $37.44_{-9.73}^{+10.71}$  & $0.10_{-0.01}^{+0.01}$ & $0.17_{-0.02}^{+0.03}$ & $0.13_{-0.04}^{+0.04}$ & $0.25_{-0.09}^{+0.11}$ & $0.10_{-0.02}^{+0.02}$ & 1.22    \\
		& 44     & 38.42   & $0.37_{-0.03}^{+0.04}$ & $2.21_{-0.86}^{+1.43}$     & $34.72_{-7.97}^{+10.17}$  & $0.11_{-0.01}^{+0.01}$ & $0.16_{-0.02}^{+0.02}$ & $0.10_{-0.04}^{+0.04}$ & $0.23_{-0.11}^{+0.13}$ & $0.08_{-0.02}^{+0.02}$ & 1.07    \\ \hline
		D7 & 45     & 8.91     & $0.62_{-0.01}^{+0.01}$ & $8.76_{-2.22}^{+4.30}$     & $5.71_{-0.17}^{+0.18}$    & $0.22_{-0.15}^{+0.14}$ & $0.43_{-0.18}^{+0.20}$ & $0.38_{-0.05}^{+0.10}$ & $0.38_{-0.12}^{+0.13}$ & $1.07_{-0.14}^{+0.13}$ & 1.33    \\
		& 46     & 16.38    & $0.71_{-0.02}^{+0.05}$ & $7.94_{-4.74}^{+44.30}$    & $4.16_{-0.86}^{+0.86}$    & $0.24_{-0.24}^{+0.23}$ & $0.25_{-0.20}^{+0.22}$ & $0.33_{-0.10}^{+0.11}$ & $0.45_{-0.16}^{+0.21}$ & $1.02_{-0.19}^{+0.29}$ & 1.33    \\
		& 47     & 22.50     & $0.80_{-0.03}^{+0.04}$ & $4.91_{-1.03}^{+3.76}$     & $4.35_{-0.25}^{+0.24}$    & $0.28_{-0.17}^{+0.16}$ & $0.05_{-0.05}^{+0.20}$ & $0.44_{-0.10}^{+0.11}$ & $0.27_{-0.10}^{+0.12}$ & $0.74_{-0.10}^{+0.14}$ & 1.13    \\
		& 48     & 27.45    & $0.82_{-0.09}^{+0.13}$ & $2.92_{-1.24}^{+3.60}$     & $3.97_{-0.70}^{+0.70}$    & $0.27_{-0.16}^{+0.14}$ & $0.20_{-0.17}^{+0.21}$ & $0.30_{-0.10}^{+0.12}$ & $0.26_{-0.11}^{+0.13}$ & $0.47_{-0.09}^{+0.13}$ & 1.15    \\
		& 49     & 31.50     & $0.41_{-0.04}^{+0.05}$ & $3.52_{-1.62}^{+2.79}$     & $34.95_{-9.32}^{+9.21}$   & $0.12_{-0.01}^{+0.02}$ & $0.23_{-0.03}^{+0.03}$ & $0.14_{-0.04}^{+0.05}$ & $0.29_{-0.05}^{+0.11}$ & $0.10_{-0.01}^{+0.02}$ & 1.32    \\
		& 50     & 36.93    & $0.40_{-0.04}^{+0.05}$ & $1.91_{-0.68}^{+1.21}$     & $32.61_{-7.81}^{+9.50}$   & $0.12_{-0.01}^{+0.01}$ & $0.22_{-0.03}^{+0.03}$ & $0.16_{-0.04}^{+0.05}$ & $0.33_{-0.10}^{+0.12}$ & $0.11_{-0.02}^{+0.03}$ & 1.39  \\  \hline
    \end{tabular}%
\\
Note: Abundances are with respect to solar \citep{Anders89}. Uncertainties are at the 90\% confidence level. The Galactic column $N_{{\rm H,Gal}}$ is fixed at $1.58\times10^{21}$ cm$^{-2}$ \citep{HI4PI16} and the LMC column $N_{{\rm H,LMC}}$ is fixed at the mean shell value $0.2\times10^{21}$ cm$^{-2}$. \\
(*) Single shock model parameters. For other regions the best-fit parameters of the ejecta component are presented.
\end{table*}

\begin{table*}[htbp]
\scriptsize
  \centering
  \caption{Comparison between this work and literature for mean shell values of DEM L71.}
    \begin{tabular}{ccccccccc}
\hline
    &    $N_{\rm H, LMC}$             &   $kT$            & $n_{\rm e}t$                 & O    & Ne  & Mg  & Si & Fe \\
    &($10^{21}$cm$^{-2}$)  &  (keV)            & ($10^{11}$cm$^{-3}$s)  &      &     &     &    &    \\
\hline
This study            & $0.20_{-0.20}^{+0.10}$ & $0.35_{-0.03}^{+0.04}$ & $ 2.47_{-1.48}^{+2.01}$ & $0.11_{-0.01}^{+0.01}$ & $0.18_{-0.02}^{+0.02}$ & $0.14_{-0.04}^{+0.05}$ & $0.34_{-0.11}^{+0.13}$ & $0.12_{-0.01}^{+0.01}$ \\
\citet{Schenck16}(**) & ---                    & $0.45_{-0.01}^{+0.01}$ & $ 2.70_{-2.10}^{+3.40}$ & $0.11_{-0.01}^{+0.01}$ & $0.20_{-0.01}^{+0.01}$ & $0.18_{-0.02}^{+0.01}$ & $0.29_{-0.03}^{+0.05}$ & $0.15_{-0.01}^{+0.01}$ \\
\citet{Hughes03}(*)  & $0.58_{-0.06}^{+0.04}$ & $0.47_{-0.03}^{+0.03}$ & $4.37_{-0.98}^{+1.25}$ & $0.21_{-0.02}^{+0.03}$ & $0.42_{-0.04}^{+0.05}$ & $0.37_{-0.07}^{+0.08}$ & $0.33_{-0.11}^{+0.11}$ & $0.09_{-0.01}^{+0.01}$ \\
\citet{Russell92}(**)  & --- & --- & --- & $0.263$ & $0.331$ & $0.316$ & $0.309$ & $0.363$ \\
\hline
    \end{tabular}%
    \\
(*) The model parameters are given for outer rim region of DEM L71. \\
(**) The ISM abundances for LMC.
\end{table*}%

\section{Discussion} 
\subsection{Nature of the Ambient Medium of DEM L71}
We studied the X-ray spectrum from the outermost shell regions to assign the features of the swept-up ISM. The mean values of our fitted model parameters for the outermost shell regions are compared with the ones given by \citet{Hughes03} and \citet{Schenck16} in Table 4. The electron temperature ($kT$) and ionization time scale ($n_{\rm e}t$) parameters obtained in this study are consistent
with the \citet{Schenck16} and \citet{Hughes03} values within statistical uncertainties. Our results show that Si and Fe abundances are consistent with values of \citet{Hughes03}. O, Ne, and Mg abundances are consistent with those by \citet{Schenck16}, which are by a factor of $\sim 2$ lower than those by \citet{Hughes03}. 

\subsection{Spatial and Chemical Structure of Ejecta}

The spectral analysis of the observed X-ray spectra for 50 sub-regions in DEM L71 shows Fe is overabundant in the central parts of the remnant up to $r\sim 22''$ (Figures 6-10). This suggests that the central metal-rich ejecta  extends roughly out to $r\sim 22''$. In contrast to Fe, the estimated abundances for O and Ne are consistent with the swept-up shell values. The average Fe/O abundance ratio of the ejecta is $\sim 10$ times higher than the solar ratio. In most azimuthal segments Si is not enhanced, and it usually reaches its maximum value at $r\sim 20''$, except D2 (west) direction (Figure 10g). While the Mg abundance is higher than the mean shell value at $r\lesssim 26''$ in almost all directions, it decreases  to  shell  value at $r\sim 12''$ in only  the west (D2). Mg  is also relatively higher in the southeastern parts (directions D6, D7) of the remnant. 

Fe is significantly enhanced, by a factor of $\sim 5-6$ on average higher than the mean shell abundances, up to $r=22''$ in the SNR for almost all azimuthal directions, however, it drops suddenly to the mean shell value at $r\sim 12''$ in the west (D2) direction (Figure 10h). Moreover, Fe is less enhanced in the east (D5) than in the north and south. It seems that Fe is primarily enhanced in the north-south direction, but not along the east-west direction. The higher density on the east and western rims of DEM L71 indicates the reverse shock moved through the ejecta on those sides more quickly, and so there should be no unshocked ejecta left. This suggest that the anisotropy really caused by a compositional asymmetry in the ejecta. Asymmetric distribution of central Fe might have been caused by an asymmetric explosion of this Type Ia SN \citep{Zhou18, Post14, Park07}. Asymmetric Type Ia explosions have been suggested by a growing number of SD models in which detonations may ignite at multiple off-center positions in the progenitor \citep[e.g.,][]{Gamezo05, Maeda10, Malone14}. Collisional DD scenarios also predict asymmetric explosions \citep{Kushnir13}. The central ejecta also shows distinctive compositions for different directions. When the western (D1, D2) and the eastern (D5) parts of the central ejecta nebula has an Fe-to-Si abundance ratio of $\sim 2$, the northern extension regions show an Fe-to-Si abundance ratio of $\sim 3.5$. The southern (directions D6, D7) part of the central ejecta nebula also has an Fe-to-Si abundance ratio of $\sim 2.5$. This may be a relic structure of the nucleosynthesis from different layers of the Si-burning \citep{Thielemann86} during the SN explosion. Si and Fe abundances are also higher than the mean shell abundances at the outer boundary of directions D3 (northwest) and D4 (northeast), respectively. Alternatively, the observed asymmetric ejecta might be the result of a spherically symmetric Type Ia SN explosion in a non-uniform circumstellar medium (CSM). A CSM-ejecta interaction can be considered in the context of a SD scenario in which the ambient medium has been modified by stellar winds from the companion or progenitor, similar to that of Kepler's SNR \citep[e.g.,][]{Chiotellis12, Patnaude12, Burkey13}. Observations show that $>$20\% of Type Ia supernovae (SNe) may be interacting with CSM released by the progenitor system prior to the explosion \citep{Sterberg11, Foley12, Maguire13}. DEM L71 may belong to this relatively small population of Type Ia SNe that interact with modified CSM. Our measured density ($n_0$ $\sim4.2$) for the western outermost boundary, which is significantly higher than that of the mean shell. This may be related to the ejecta distribution on the west part of the remnant, and thus, the observed ejecta distribution toward the west might be due to this ambient density gradient rather than the intrinsically asymmetric explosion, or both.

In general, the composition of the metal-rich ejecta shows enhanced Fe abundances, with a lack of O, in general agreement with the previously identified Type Ia origin for this SNR. Furthermore the Fe-rich central ejecta are hotter ($r<25''$, $kT\sim 0.75$ keV) than the outer parts ($r>25''$, $kT\sim 0.40$ keV) as shown in Figure 10a. Emission measure is compatible with surface brightness distribution and generally peaks $r\sim 30''$ (Figure 10b). This is because the emission from the swept-up ISM peaks at this angular distance. The ionization timescale of the metal-rich ejecta gas in DEM L71 generally has similar radial profiles and it is high on the central parts of the remnant. It also peaks at $\sim 28''- 30''$ from the SNR's X-ray geometric center and decreases beyond $r\sim 30''$ (Figure 10c). Forward-shocked swept-up ISM dominates toward the outer boundary $r\geq 30''$. Fe abundances generally decrease towards the mean shell value at $r\sim 22''$. This suggests that the location of the contact discontinuity (CD) which separates the shocked supernova ejecta and shocked ISM at $r\sim 22''$ generally corresponding to $\sim 5$ pc from the geometric center of SNR at a distance of 50 kpc to the LMC \citep{Freedman01} (Figure 11). Besides that location of CD is complicated and uncertain in the west direction. In Figure 11, the circle on three colour image of the remnant is consistent with the boundary of the bluish central nebula, but the western part of the circle include regions with significantly different colours. 

We compare our results for regions 38 and 39 with the values given by \citet{Hughes03}, who examined the similar regions of DEM L71. The O, Ne, Mg, and Si abundances calculated in our study were $\sim$ 2-5 times lower than the \citet{Hughes03} values, while the Fe abundances were generally consistent within statistical uncertainties (see Table 5). There is also an inconsistency between our $kT$ and $n_{\rm e}t$ measurements and the \citet{Hughes03} values. Plasma parameter and elemental abundance measurements are model-dependent. Our model utilizes improved atomic data compared to what \citet{Hughes03} used. The reason why temperatures are lower in new analyses may be due to the fact that current models contain more lines. Our data (a total of $\sim$100 ks exposure of the {\it Chandra} ACIS) are also a factor of $\sim$2 deeper than those used by \citet[][45.4 ks exposure of the {\it Chandra} ACIS]{Hughes03} providing a significantly higher $S/N$. These improved photon count statistics allow us more accurate estimates of elemental abundances and plasma parameters. In order to investigate the dependence of these discrepancies on the exposure time, we extracted spectra of 38 and 39 regions from the OBSID 775 data (45.4 ks) used by \citet{Hughes03}. Then, we analyzed the spectra of these regions with current plane shock models (see Section 3) and obtained results consistent with our findings in Table 5, within uncertainties (see Model 1 in Table 5). There was a discrepancy between the calculated $N_{\rm H, LMC}$ values. It was also remarkable that the $N_{\rm H, GAL}$ used by \citet{Hughes03} in planar-shock model was $\sim$3 times lower than ours. To test whether this caused a parameter degeneration, the models were reapplied with \citet{Hughes03}'s $N_{\rm H}$ values (see Model 2 in Table 5). Considering the results in these models made with \citet{Hughes03}'s $N_{\rm H}$ values, it is seen that the elemental abundances generally consistent with the others within uncertainties while the $kT$ values are higher and the $n_{\rm e}t$ values are smaller. This systematic difference between $kT$ and $n_{\rm e}t$ points to a parameter degeneration resulting from the different $N_{\rm H}$ values in the two studies. 
 In conclusion, the discrepancies between \citet{Hughes03} and our findings may caused by the combined effect of the models, the exposure time, and the $N_{\rm H}$ values used in the spectral models. The increase in exposure time for DEM L71, the development of spectral models over time, and the use of more sensitively determined $N_{\rm H, GAL}$ in the models indicate that the findings in this study are more precise.

\begin{table*}
\scriptsize
  \centering
  \caption{Comparison between this work and \citet{Hughes03} for ejecta values of DEM L71.}
    \begin{tabular}{ccccccccccc}
\hline
    &   Data              &   $N_{\rm H,GAL}$             &  $N_{\rm H,LMC}$             &   $kT$            & $n_{\rm e}t$                 & O    & Ne  & Mg  & Si & Fe \\
   & (ks) & ($10^{21}$cm$^{-2}$) & ($10^{21}$cm$^{-2}$)  &  (keV)            & ($10^{11}$cm$^{-3}$s)  &      &     &     &    &    \\
\hline
\hline
This study (Region 38) & $100$ &  1.60$^{\rm a}$    & 0.20 & $0.64_{-0.01}^{+0.01}$ & $ 9.67_{-2.01}^{+3.41}$ & $0.17_{-0.05}^{+0.05}$ & $0.14_{-0.11}^{+0.12}$ & $0.25_{-0.06}^{+0.06}$ & $0.30_{-0.09}^{+0.09}$ & $1.03_{-0.11}^{+0.14}$ \\
\citet{Hughes03}  (inner core) & $45.4$  & 0.56$^{\rm b}$  & $0.52_{-0.14}^{+0.24}$ & $1.10_{-0.30}^{+0.30}$ & $1.00_{-0.31}^{+1.04}$ & $0.33_{-0.09}^{+0.27}$ & $1.70_{-0.80}^{+1.50}$ & $1.10_{-0.50}^{+1.20}$ & $1.00_{-0.50}^{+1.10}$ & $1.70_{-0.60}^{+1.20}$ \\
\hline
This study (Region 39)   & $100$ &     1.60$^{\rm a}$  & 0.20 & $0.74_{-0.02}^{+0.01}$ & $ 4.24_{-0.61}^{+0.82}$ & $0.05_{-0.05}^{+0.08}$ & $0.04_{-0.04}^{+0.19}$ & $0.37_{-0.08}^{+0.08}$ & $0.42_{-0.09}^{+0.09}$ & $1.05_{-0.11}^{+0.13}$ \\
\citet{Hughes03}  (outer core)  & $45.4$  & 0.56$^{\rm b}$  & $0.67_{-0.17}^{+0.31}$ & $1.60_{-0.50}^{+0.70}$ & $0.63_{-0.19}^{+0.42}$ & $0.31_{-0.09}^{+0.25}$ & $2.30_{-0.90}^{+1.90}$ & $1.70_{-0.70}^{+1.60}$ & $0.70_{-0.50}^{+1.00}$ & $1.90_{-0.60}^{+1.40}$ \\
\hline
Model 1 (Region 38)   & $45.4$ & 1.60$^{\rm a}$    & 0.20 & $0.67_{-0.04}^{+0.07}$ & $ 3.64_{-1.74}^{+8.68}$ & $0.18_{-0.07}^{+0.07}$ & $0.02_{-0.02}^{+0.13}$ & $0.20_{-0.08}^{+0.08}$ & $0.41_{-0.17}^{+0.25}$ & $1.22_{-0.26}^{+0.45}$ \\
Model 2 (Region 38)  & $45.4$ & 0.56$^{\rm b}$  & 0.52$^{\rm c}$ & $1.03_{-0.04}^{+0.03}$ & $0.72_{-0.16}^{+0.52}$ & $0.16_{-0.02}^{+0.02}$ & $0.05_{-0.02}^{+0.12}$ & $0.31_{-0.10}^{+0.10}$ & $0.63_{-0.18}^{+0.19}$ & $1.85_{-0.14}^{+0.16}$ \\
\hline
Model 1 (Region 39)      & $45.4$ &   1.60$^{\rm a}$  & 0.20 & $0.72_{-0.01}^{+0.01}$ & $ 9.98_{-1.88}^{+10.18}$ & $0.03_{-0.03}^{+0.17}$ & $0.03_{-0.02}^{+0.32}$ & $0.42_{-0.12}^{+0.12}$ & $0.42_{-0.14}^{+0.14}$ & $1.09_{-0.06}^{+0.21}$ \\
Model 2 (Region 39)     & $45.4$  &  0.56$^{\rm b}$  & 0.67$^{\rm c}$ & $1.18_{-0.14}^{+0.12}$ & $ 0.88_{-0.14}^{+0.32}$ & $0.08_{-0.02}^{+0.02}$ & $0.03_{-0.03}^{+0.11}$ & $0.28_{-0.08}^{+0.08}$ & $0.32_{-0.10}^{+0.15}$ & $1.04_{-0.08}^{+0.26}$ \\
\hline
\hline
    \end{tabular}%
\\
Note: $^{\rm a}$ from \citet{HI4PI16}, $^{\rm b}$ from \citet{Dickey90}, $^{\rm c}$ from  \citet{Hughes03}\\ 
\end{table*}%

\begin{figure*}
\begin{center}
\includegraphics[scale=0.45]{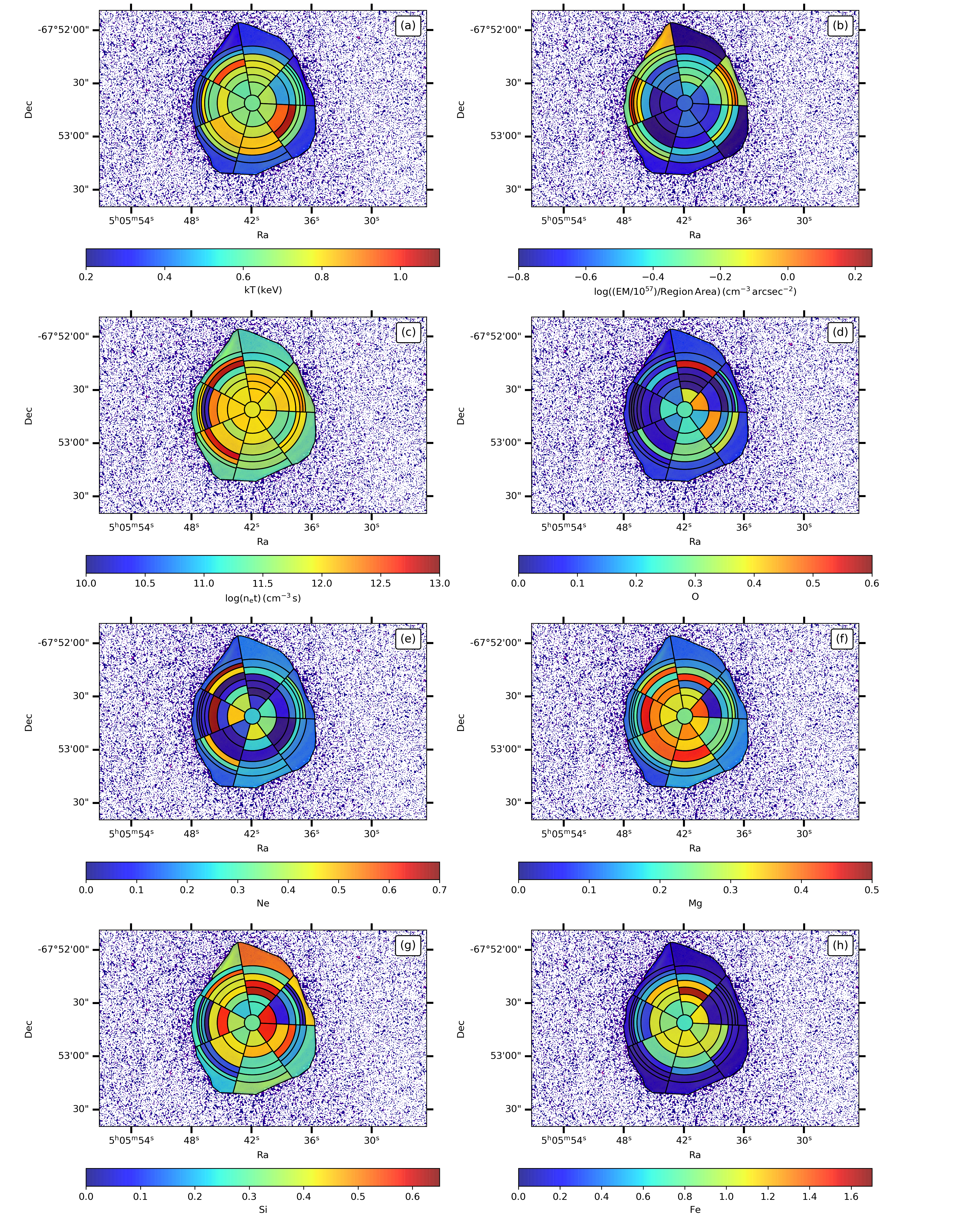}
\caption{The plasma parameters and elemental abundances distribution on DEM L71. Electron temperatures $kT$  (a), emission measures $EM$ (b), ionization timescales $n_{\rm e}t$ (c), O abundance (d), Ne abundance (e) Mg abundance (f), Si abundance (g) and Fe abundance (h). Abundances are with respect to solar \citep{Anders89}.}
\end{center}
\end{figure*}

\begin{figure}
\begin{center}
\includegraphics[scale=.9]{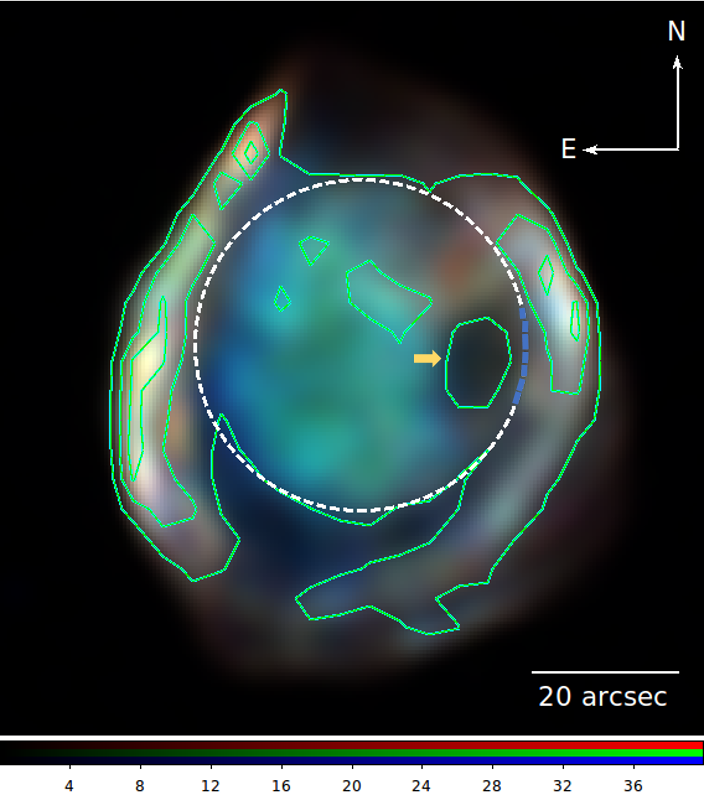}
\caption{The estimated position of the CD is represented by the dashed-white circle on the three-colour image of DEM L71.
The yellow arrow points to the region with the low Fe abundance in the central part of the remnant. The dashed-blue part of the circle represents uncertain CD location. The broadband X-ray image contours of the SNR are overlaid.} 
\end{center}
\end{figure}

\subsection{SNR Dynamics}

To estimate the explosion energy and the age of the SNR, we apply self-similar Sedov solutions \citep{Sedov59}. For these purposes, based on the volume emission measure values ($EM = n_{\rm e}n_{\rm H}V$) estimated from the best-fit spectral models of the shell regions we calculate the post-shock electron density ($n_{\rm e}$). For this estimation we calculated the X-ray emitting volumes ($V$) for each region are listed in Table 6. All shell regions we also assumed for a $\sim 1.2$ pc path length (roughly corresponding to the angular thickness of the each shell region at 50 kpc) along the line of sight. For a mean charge state with normal composition, we assumed $n_{\rm e}\sim 1.2n_{\rm H}$ (where $n_{\rm H}$ is the H number density) and calculated electron density for all shell region $n_{\rm e} \sim 8.2-20.4 f^{-1/2}{\rm cm}^{-3}$ where $f$ is the volume filling factor of the X-ray emitting gas. The pre-shock H density $n_0$ are listed in Table 6 assuming a strong adiabatic shock where $n_{\rm H}=4n_0$. Under the assumption of electron-ion temperature equipartition for DEM L71 \citep{Ghavamian07}, the gas temperature is related to the shock velocity ($V_{\rm s}$) as $T=3\hat{m}v_{\rm s}^2/16k$ (where $\hat{m}\sim0.6m_{\rm p}$ and $m_{\rm p}$ is the proton mass). Using electron temperatures, we calculated shock velocity of $V_{\rm s}\sim 472-603$ km s$^{-1}$ and Sedov age of $\tau_{\rm sed}\sim 5,568-8,142$ yr for each shell region (see Table 6). The median shock velocity and Sedov age calculated from the eight shell region values are $\sim 545\pm 37$ km s$^{-1}$ and $\sim 6,660\pm 770$ yr, respectively. Fe which primarily enhanced in the north-south direction, but not along the east-west direction, indicates that the material in the north-south direction is more thoroughly burned to Fe, and thus more energy is released along that direction. \citet{Ghavamian03} and \citet{Rakowski03} revealed that the shock velocities in the east direction of the remnant are lower than in the other directions, and this supports our proposal. Considering that the remnant expands further in the north-south direction, it is possible that the initial shock velocity is higher in this direction. Our shock-velocity estimate is only a conservative lower limit, therefore our SNR age estimate is an upper limit. Although our age upper limit is not tightly constraining, it is generally consistent with previous estimates of $4,360\pm290$ yr \citep{Ghavamian03} and $\sim5,000$ yr \citep{Hughes98}. We calculated the corresponding explosion energy of $E_0\sim 0.96-2.00\times 10^{51}$ erg for DEM L71 (see Table 6). The median explosion energy was $E_0=1.74\pm 0.35 \times 10^{51}$ erg which is consonant with the canonical $E_0\ = 10^{51}$ erg. In the calculations of $\tau_{\rm sed}$ and $E_0$, the distance of each shell region ($7.9-10.7$ pc) from the geometric center of the remnant is based. The total swept-up mass, $M_{\rm SW}=n_0m_{\rm p}V$ is also estimated to be $M_{\rm SW} \sim 145_{-25}^{+30} $ $f^{1/2}M_{\odot}$, assuming that the average SNR radius of $\sim 9.7$ pc. This calculated mass conforms with the swept-up ISM value given by \citet{Siegel20} within uncertainties. Our estimated explosion energy  suggest that DEM L71 originates from a canonical SD or DD progenitor system. 

\begin{table*}
\scriptsize
  \centering
  \caption{Assumed $V$, $n_0$ values for Sedov solutions, and derived dynamic parameters ($V_{\rm s}$, $\tau_{\rm sed}$, and $E_0$)  of DEM L71.}
    \begin{tabular}{cccccc}
\hline
Shell Region& $V$ & $n_0$        & $V_{\rm s}$        & $\tau_{\rm sed}$ & $E_0$\\
& (10$^{56}$ cm$^{3}$) &  (cm$^{-3}$) &  (km s$^{-1})$ & (yrs)  &  ($\times 10^{51}$erg) \\        
\hline
S1	& 3.75	& 1.94	& 512 & 8,142 & 1.90\\
S2	& 1.41	& 4.24	& 472 & 6,833 & 1.62\\
S3	& 4.14	& 1.71	& 534 & 7,097 & 1.37\\
S4	& 1.42	& 2.12	& 603 & 5,974 & 1.85\\
S5	& 2.83	& 1.85	& 558 & 7,309 & 2.00\\
S6  & 1.42  & 3.02  & 553 & 6,173 & 1.88\\ 
S7	& 2.16	& 2.09	& 553 & 5,568 & 0.96\\
S8	& 2.11	& 2.63	& 508 & 6,478 & 1.24\\

\hline
    \end{tabular}%
\end{table*}%

\section{Summary \& Conclusion}
We present the results of our extensive analysis of the {\it Chandra} archival data of Type Ia SNR DEM L71 in the LMC. Our detailed spatially-resolved spectral analysis of the entire remnant reveals the complete spatial distribution of the shocked metal-rich ejecta in DEM L71. The ejecta material is mainly contained within a $\sim 10$ pc diameter circular region at the geometric center of the remnant. We also detect an asymmetric structure of central metal-rich ejecta material in the western part of the remnant. The asymmetric distribution of metal-rich ejecta gas is likely caused by an asymmetric explosion of the progenitor, but it should not be ruled out that the ejecta may undergo in a non-uniform expansion in interstellar material with different densities. We estimate the explosion energy $E_0\sim 1.74\pm 0.35\times 10^{51}$ erg. This explosion energy estimate is compatible with a canonical explosion of a Type Ia SNR. We also estimate a Sedov age of $\sim 6,660\pm 770$ yr for DEM L71. Our estimated value for the explosion energy consistent with a canonical explosion of a Type Ia supernova remnant.

\section*{Acknowledgements}
We thank the anonymous referee for his/her insightful and constructive suggestions, which significantly improved the paper. This study was funded by Scientific Research Projects Coordination Unit of Istanbul University. Project number: FOA-2018-30716. We are grateful to Dr. Sangwook Park for the inspiration and helpful discussions. Also, we would like to thank Olcay Plevne, O\v{g}uz Han Ata\c{s} and Dr. Jayant Bhalerao for their contributions. This research has made use of data obtained from the Chandra Data Archive and the Chandra Source Catalog, and software provided by the Chandra X-ray Center (CXC) in the application packages CIAO, ChIPS, and Sherpa. This research has made use of NASA's Astrophysics Data System Bibliographic Services.

\section*{Data Availability}
The X-ray data on DEM L71 as described in Section 2 include  Chandra ACIS-S observations and data are available in the Chandra archive (https://asc.harvard.edu/cda/). Processed data products underlying this article will be shared on reasonable request to the authors.





\begin{thebibliography}{99}
\bibitem[Anders \& Grevesse(1989)]{Anders89} 
Anders E., Grevesse N., 1989, Geochimica et Cosmochimica Acta, 53, 197 

\bibitem[Badenes et al.(2006)]{Badenes06}
Badenes C., Borkowski K. J., Hughes J. P., Hwang U., Bravo E., 2006, ApJ, 645, 1373 

\bibitem[Bautz et al.(1998)]{Bautz98} 
Bautz M. W., Pivovaroff M., Baganoff F. et al., 1998, SPIE, 3444, 210

\bibitem[Borkowski et al.(2001)]{Borkowski01} 
Borkowski K. J., Lyerly W. J., Reynolds S. P., 2001, ApJ, 548, 820 

\bibitem[Burkey et al.(2013)]{Burkey13} 
Burkey M., Reynolds S., Borkowski K., et al., 2013, ApJ, 764, 63

\bibitem[Chiotellis et al.(2012)]{Chiotellis12}
Chiotellis A., Schure K., Vink J., 2012, A\&A, 537, 139

\bibitem[Davies et al.(1976)]{Davies76}
Davies R. D., Elliott K. H., Meaburn J., 1976, MmRAS, 81, 89

\bibitem[Dickey \& Lockman(1990)]{Dickey90}
Dickey J. M., Lockman F. J. 1990, ARA\&A, 28, 215

\bibitem[Dopita et al.(1981)]{Dopita81}
Dopita M. A., Tuohy I. R., Mathewson D. S., 1981, ApJ, 248, L105 

\bibitem[Foley et al.(2012)]{Foley12} Foley R.~J., Simon J.~D., Burns C.~R. et al., 2012, ApJ, 752, 101

\bibitem[Foster et al.(2012)]{Foster12}
Foster A. R., Ji L., Smith R. K., Brickhouse N. S., 2012, ApJ, 756, 128

\bibitem[Freedman et al.(2001)]{Freedman01}
Freedman W. L., Madore B. F., Gibson B. K. et al., 2001, ApJ, 553, 47

\bibitem[Fruscione et al.(2006)]{Fruscione06}
Fruscione A., McDowell J.~C., Allen G.~E., Brickhouse N.~S., Burke D.~J., Davis J.~E., Durham N., et al., 2006, SPIE, 6270

\bibitem[Gamezo, Khokhlov, \& Oran(2005)]{Gamezo05}
Gamezo V.~N., Khokhlov A.~M., Oran E.~S., 2005, ApJ, 623, 337

\bibitem[Ghavamian et al.(2003)]{Ghavamian03}
Ghavamian P., Rakowski C. E., Hughes J. P., Williams T. B., 2003, ApJ, 590, 833

\bibitem[Ghavamian et al.(2007)]{Ghavamian07}
Ghavamian P., Laming J. M., Rakowski, C. E., 2007, ApJL, 654, L69

\bibitem[HI4PI Collaboration et al.(2016)]{HI4PI16} 
HI4PI Collaboration, Ben Bekhti N., Floer L., et al., 2016, A\&A, 594, A116

\bibitem[Hughes et al.(1998)]{Hughes98} 
Hughes J. P., Hayashi I., Koyama K., 1998, ApJ, 505, 732

\bibitem[Hughes et al.(2003)]{Hughes03} 
Hughes J. P., Ghavamian P., Rakowski C. E., Slane, P. O., 2003, ApJ, 582L, 95

\bibitem[Kushnir et al.(2013)]{Kushnir13} 
Kushnir D., Katz B., Dong S., Livne E., Fern{\'a}ndez R., 2013, ApJL, 778, L37

\bibitem[Long et al.(1981)]{Long81} 
Long K. S., Helfand D. J., Grabelsky D. A., 1981, ApJ, 248, 925

\bibitem[Iben \& Tutukov(1984)]{Iben84} 
Iben I. Jr., Tutukov A. V., 1984, ApJ, 282, 615

\bibitem[Maeda et al.(2010)]{Maeda10}
Maeda K., Benetti S., Stritzinger M., et al., 2010, Natur, 466, 82

\bibitem[Maggi et al.(2016)]{Maggi16}
Maggi P., Haberl F., Kavanagh P. J. et al., 2016, A\&A, 585A, 162

\bibitem[Maguire et al.(2013)]{Maguire13} Maguire K., Sullivan M., Patat F., et al., 2013, MNRAS, 436, 222

\bibitem[Malone et al.(2014)]{Malone14}
Malone C.~M., Nonaka A., Woosley S.~E., Almgren A.~S., Bell J.~B., Dong S., Zingale M., 2014, ApJ, 782, 11

\bibitem[Maoz et al.(2014)]{Maoz14} 
Maoz D., Mannucci F., Nelemans G., 2014, ARA\&A, 52, 107

\bibitem[McConnachie(2012)]{McConnachie12}
McConnachie A. W., 2012, AJ, 144, 4

\bibitem[Nomoto(1982)]{Nomoto82} 
Nomoto K., 1982, ApJ, 253, 798

\bibitem[Pagnotta \& Schaefer(2015)]{Pagnotta15}
Pagnotta A, Schaefer B. E., 2015, ApJ, 799, 101

\bibitem[Park et al.(2006)]{Park06} 
Park T., Kashyap V. L., Siemiginowska A., van Dyk D. A., Zezas A., Heinke C., Wargelin B. J.,  2006, ApJ, 652, 610 

\bibitem[Park et al.(2007)]{Park07} 
Park S., Slane P.~O., Hughes J.~P., Mori K., Burrows D.~N., Garmire G.~P., 2007, ApJ, 665, 1173

\bibitem[Patnaude et al.(2012)]{Patnaude12} 
Patnaude D., Badenes C., Park S., Laming J., 2012, ApJ, 756, 6

\bibitem[Post et al.(2014)]{Post14} 
Post S., Park S., Badenes C., Burrows D.~N., Hughes J.~P., Lee J.-J., Mori K., et al., 2014, ApJL, 792, L20

\bibitem[Rakowski et al.(2003)]{Rakowski03} 
Rakowski C. E., Ghavamian P., Hughes J. P., 2003, ApJ, 590, 846

\bibitem[Ruiz-Lapuente(2014)]{Ruiz-Lapuente14} 
Ruiz-Lapuente P., 2014, NewAR, 62, 15

\bibitem[Ruiz-Lapuente et al.(2018)]{Ruiz-Lapuente18} 
Ruiz-Lapuente P., Damiani F., Bedin L. et al., 2018, ApJ, 862, 124 

\bibitem[Russell \& Dopita(1992)]{Russell92}
Russell S. C., Dopita M. A., 1992, ApJ, 384, 508 

\bibitem[Schenck et al.(2014)]{Schenck14}
Schenck A., Park S., Burrows D. N., Hughes J. P., Lee J., Mori K., 2014, ApJ, 791, 50

\bibitem[Schenck et al.(2016)]{Schenck16}
Schenck A., Park S., Post S., 2016, AJ, 151, 161 

\bibitem[Sedov(1959)]{Sedov59}
Sedov, L.~I., 1959, Similarity and Dimensional Methods in Mechanics, New York: Academic Press

\bibitem[Siegel et al.(2020)]{Siegel20}
Siegel J., Dwarkadas V. V., Frank K., Burrows D. N., Panfichi A., 2020, AN, 341, 163

\bibitem[Smithsonian Astrophysical Observatory(2000)]{ds9cite}
Smithsonian Astrophysical Observatory. 2000, SAOImageDS9: A utility for displaying astronomical images in theX11 window environment.  http://ascl.net/0003.002

\bibitem[Sterberg et al.(2011)]{Sterberg11}
Sterberg A., Gal-Yam A., Simon D., et al., 2011, Science, 333, 856

\bibitem[Thielemann et al.(1986)]{Thielemann86}
Thielemann F., Nomoto K., Yokoi K., 1986, A\&A, 158, 17
 
\bibitem[van der Heyden et al.(2003)]{vanderHeyden03}
van der Heyden K. J., Bleeker J. A. M., Kaastra J. S., Vink J., 2003, A\&A, 406, 141

\bibitem[Wang \& Han(2012)]{Wang12}
Wang B., Han Z., 2012, NewAR, 56, 122

\bibitem[Wang(2018)]{Wang18}
Wang B, 2018, RAA, 18, 49

\bibitem[Webbink(1984)]{Webbink84} 
Webbink R. F., 1984, ApJ, 277, 355

\bibitem[Whelan \& Iben(1973)]{Whelan73} 
Whelan J., Iben I., 1973, ApJ, 186, 1007

\bibitem[Zhou \& Vink(2018)]{Zhou18} 
Zhou P.,  Vink J., 2018, A\&A, 615, A150, 14

\end{thebibliography}

\bsp	
\label{lastpage}
\end{document}